\documentclass{elsart}
\usepackage{graphicx,color}
\usepackage{amsmath}

\usepackage{amsmath}
\usepackage{amssymb}

\def\nab{{\boldsymbol \nabla}}

\def\simle{\mathrel{\rlap{\raise 0.511ex \hbox{$<$}}{\lower 0.511ex \hbox{$\sim$}}}}
\def\simge{\mathrel{ \rlap{\raise 0.511ex \hbox{$>$}}{\lower 0.511ex \hbox{$\sim$}}}}
 
\def\q{{\boldsymbol q}}

\newcommand \beq{\begin{eqnarray}}
\newcommand \eeq{\end{eqnarray}} 
\def\simle{\mathrel{\rlap{\raise 0.511ex \hbox{$<$}}{\lower 0.511ex 
\hbox{$\sim$}}}}
\def\simge{\mathrel{ \rlap{\raise 0.511ex 
\hbox{$>$}}{\lower 0.511ex \hbox{$\sim$}}}}
\newcommand{\del}{\partial}

\def\x{{\boldsymbol x}}
\def\v{{\boldsymbol v}}

\def\q{{\boldsymbol q}}
\def\p{{\boldsymbol p}}

\begin{document}
\begin{frontmatter}
\title{Gluon Transport Equation in the Small Angle Approximation and the Onset of Bose-Einstein Condensation}
\author[cea]{Jean-Paul Blaizot}
\author[ind,rbrc]{Jinfeng Liao}
\author[rbrc,bnl,ccnu]{Larry McLerran}

\address[cea]{Institut de Physique Th\'eorique, CNRS/URA 2306, CEA Saclay,
F-91191 Gif-sur-Yvette, France}
\address[ind]{Physics Department and Center for Exploration of Energy and Matter,
Indiana University, 2401 N Milo B. Sampson Lane, Bloomington, IN 47408, USA}
\address[rbrc]{RIKEN BNL Research Center, Bldg. 510A, Brookhaven National Laboratory,
   Upton, NY 11973, USA}
   \address[bnl]{Physics Dept, Bdg. 510A, Brookhaven National Laboratory, Upton, NY-11973, USA}
   \address[ccnu]{Physics Department, China Central Normal University, Wuhan, China}

\begin{abstract}
 In this paper, we study the evolution of a  dense system of gluons, such as those produced in the early stages of ultra-relativistic heavy ion collisions. We describe the approach to thermal equilibrium using  the small angle approximation  for gluon scattering  in a Boltzmann equation that  includes the effects of Bose statistics. In the present study we ignore the effect of the longitudinal expansion, i.e., we restrict ourselves to spatially uniform systems, with spherically symmetric momentum distributions. Furthermore we take into account only elastic scattering, i.e., we neglect  inelastic, number changing, processes.  We solve the transport equation for various initial conditions that correspond to small or large initial gluon phase-space densities. For a small initial phase-space density,  the system evolves towards  thermal equilibrium, as expected. For a large enough initial phase-space density the equilibrium state contains a Bose condensate. We present numerical evidence that such over-populated systems reach the onset of Bose-Einstein condensation in a finite time. The  approach to  condensation is characterized by a scaling behavior that we briefly analyze.
\end{abstract}
 \end{frontmatter}

\section{Introduction}

It has been argued recently \cite{Blaizot:2011xf} that the gluonic systems created in the early stages of nucleus-nucleus collisions contain more gluons than can be accommodated by a Bose-Einstein equilibrium distribution. More precisely, the dimensionless number $ n/\epsilon^{3/4} $, with $n$ the number density and $\epsilon$ the energy density, 
exceeds typically the maximal value allowed by a thermally equilibrated Bose distribution function. We shall refer to such systems as ``over-occupied''. For initial conditions where all modes are occupied uniformly up to the saturation scale $Q_s$, with occupation factor of order $1/\alpha_s$, the over-occupation parameter is $n/\epsilon^{3/4} \sim 1/\alpha^{1/4}$. It can therefore be  large when the coupling constant $\alpha_s$ is small\footnote{In fact, we shall see in the next section (see after Eq.~(\ref{overpopparam})) that,  for the relevant initial conditions, over-occupation occurs even when the coupling constant is not very small.}. Under such conditions, 
if the kinetic of the approach to equilibrium is dominated by processes that conserve the number of particles,  a Bose condensate will develop as the system evolves towards equilibrium. 

In this paper, we study this phenomenon using a Boltzmann equation for gluons in the small angle scattering limit. The virtue of the small angle approximation is that it converts the transport equation into a Fokker-Planck equation that is much easier to solve. Such an equation has already been used to study some features of the thermalization of the quark-gluon plasma \cite{Mueller:1999pi}. However, both in \cite{Mueller:1999pi}, and in the numerical study presented in Ref.~\cite{Bjoraker:2000cf}, the  quantum Boltzmann equation was treated only in the regime of small occupations, and the possible amplification of soft modes by Bose statistical factors was not considered. It is assumed in these works that the plasma frequency provides an infrared cutoff on the energy of the soft modes, that prevents the occupation factor to grow as the momentum becomes vanishingly small. However,  if elastic scatterings play a dominant role, one may expect a chemical potential to quickly develop. This will eventually cancel the effect of the plasma frequency, thereby  allowing the unlimited growth of the soft modes that can drive the system towards condensation.  

 The nature of such a condensate in non abelian gauge theories raises a number of conceptual issues that will not be addressed here. In fact, in QCD, inelastic processes eventually preclude the existence of a condensate in equilibrium.  Whether a transient condensate forms dynamically is an interesting open issue.  Our present study reveals a robust tendency for the rapid growth of soft modes, enhanced by the Bose statistical factors. A recent study of the effect of inelastic scatterings within the same context \cite{Huang:2013lia} suggests that this rapid growth is not tamed by number changing processes, but on the contrary may  be amplified. The issue of BEC formation has also been addressed in the context of cosmology \cite{Semikoz:1995rd}. There, the relevant dynamics is that of a scalar field, and the condensate has a natural definition in terms of the expectation value of the scalar field. Evidence for the formation of a transient condensate has  been obtained in classical statistical field simulations of scalar fields \cite{Berges:2012us,Dusling:2010rm}. Classical statistical field simulations have also been performed for gauge fields with,  however, no clear-cut conclusion yet (see e.g \cite{Berges:2013eia}, and references therein). Clearly more work is needed before we have a clear picture of how important this phenomenon is for the  thermalization of the quark-gluon plasma in ultra-relativistic heavy ion collisions.\footnote{For a recent discussion of some of the issues involved in the thermalization of the quark-gluon plasma, see \cite{Berges:2012ks} and references therein.} 

But aside from its relevance to the thermalization of the quark-gluon plasma, the kinetics of condensation is an interesting problem in itself. Thus, in the present paper, we shall  leave aside many important aspects of  the dynamics of a quark-gluon plasma at weak coupling \footnote{For a thorough discussion of the many aspects of this problem see \cite{Kurkela:2011ub}.}, and focus on an idealized situation, that of  a static system of gluons in a box, with an isotropic momentum distribution, and study how such systems evolve towards equilibrium, starting from over-occupied initial conditions. Analogous problems have been addressed in other contexts. Close to our study are the works of Ref.~\cite{Semikoz:1995rd} for the massive scalar field,  and that of Ref.~\cite{Lacaze:2001qf} for the non relativistic atomic gases. We depart from those studies in that we consider a system of gluons, instead of scalar particles or non-relativistic atoms. We solve the Boltzmann equation in the small angle approximation, instead of the full Boltzmann equation; this approximation is legitimate in QCD where the dominant scatterings are at small angles. And finally we consider massless particles, so that the dispersion relation is ultra-relativistic instead of non relativistic as it is in the works just quoted. However, as in these works,  we find that the onset of condensation is preceded by a scaling regime, which we briefly analyze. Note that the kinetic equation that we used in this paper allows us to describe the system only up to the onset of condensation. We defer to a later publication the discussion of the kinetics in the presence of a fully developed condensate.

The plan of this paper is as follows. In the next section we review briefly the properties of the transport equation in the small angle approximation. In the following section, we discuss the two types of solutions that are expected, depending upon whether the system is initially under or over populated: in the first case, the system evolves towards a Bose-Einstein equilibrium distribution at large times; in the second case, the onset of Bose-Einstein condensation is reached in a finite time. These two cases are studied numerically in Sects. 4 and 5, respectively. Sect. 6 summarizes the conclusions and points out natural extensions of the present study. 

\section{The Transport  Equation in the Small Angle Approximation}

We  describe the system of gluons that is produced in an ultra-relativistic heavy ion collision in terms of a distribution function in phase-space, 
\beq
    f(\x,\p)=\frac{(2\pi)^3}{2(N_c^2-1)}\frac{dN}{d^3\x d^3\p},
    \eeq
where $N$ denotes the total number of gluons in the system.  In other words, $f$ denotes the number of gluons of a given spin and color in the phase-space element $d^3\x d^3\p/(2\pi)^3$. We  assume that $N$ remains constant during the evolution, that is, we ignore in this work the effect of inelastic, number changing, processes. We also assume that the phase-space densities of different spins and colors are equal, that is, $f$ is independent of spin and color. 

The kinetic equation that describes the evolution of  bosons  by elastic scattering and that  includes the effect of Bose statistics (sometimes referred to as the Boltzmann-Nordheim equation),  is of the form:
\begin{eqnarray} \label{Eq_transport}
 && {\mathcal D}_t f_1=  \frac{1}{2} \int {{d^3p_2} \over {(2\pi)^3 2 E_2}}{{d^3p_3} \over {(2\pi)^3 2 E_3}}
   {{d^3p_4} \over {(2\pi)^3 2 E_4}} {1 \over {2E_1}}  \mid M_{12\to 34} \mid^2  \nonumber \\
 & &\times (2\pi)^4 \delta(p_1 + p_2 - p_3 -p_4)  \{f_3f_4(1+f_1)(1+f_2) -  f_1f_2(1+f_3)(1+f_4)  \} \nonumber \\
  \end{eqnarray}
    where the factor $1/2$ in front of the integral is a symmetry factor (that takes into account the fact that each configuration of the gluons 3 and 4 is counted twice in the sum-integrations over momenta, colors and spins). Summation over color and polarization is performed on the gluons 2,3,4; average over color and polarization is performed for gluon 1.  The `time-derivative' ${\mathcal D}_t $ on the left hand side may take into account the possible longitudinal expansion of the system, that is 
    $
    {\mathcal D}_t \equiv \partial_t + \v_z \cdot \nab,$ with $\nab$ a spatial gradient. However, in this paper we restrict ourselves to the study of a uniform, non-expanding system and ${\mathcal D}_t =\del_t$ simply.

 The matrix element for gluon-gluon scattering ($1+2\to 3+4$) summed over spin and color of 2, 3 and 4, and averaged over spin and color of 1,  is equal to (see e.g. \cite{Peskin:1995ev})
 \begin{eqnarray}
 \mid M_{12\to 34} \mid^2 = 72g^4 \left[ 3- \frac{t\, u}{s^2} -  \frac{s\, u}{t^2} - \frac{t\, s}{u^2} \right],
 \eeq
 with $s,t,u$ the usual  Mandelstam variables $ s=(p_1+p_2)^2$, $ t=(p_1-p_3)^2$, $ u=(p_1-p_4)^2$, and $p_i=(E_i, \p_i)$. The dominant scatterings are small angle scatterings, involving a small momentum transfer between the colliding gluons. The matrix element is then dominated by the $t$ and $u$ channels, and can be simplified.

 Under the assumption that small angle scatterings dominate, one can in fact reduce the transport equation to a  Fokker-Planck equation describing diffusion in momentum space \cite{LifshitzPitaevskii}.  This is achieved by writing the collision integral as the divergence of a current ${\mathcal J} (\p)$ in momentum space, 
    \beq
    {\mathcal D}_t f=-\nab\cdot {\mathcal J}=-\frac{\del{\mathcal J}_i}{\del p_i}.
\eeq
A standard calculation yields \cite{LifshitzPitaevskii}
\beq \label{current0}
{\mathcal J}_i = \frac{9}{4\pi}g^4 {\mathcal L}  \int {{d^3p_2} \over {(2\pi)^3}}\,
\left[ h_{\p_1}\nabla^j f_{\p_2} - h_{\p_2} \nabla^j f_{\p_1}  \right]{\mathcal V}_{i\, j},
\eeq
with
\begin{eqnarray}
{\mathcal V}^{i\, j} \equiv \delta^{i j} (1- \v_1\cdot\v_2) + \left( v_1^i v_2^ j + v_1^ j v_2^ i \right),
\end{eqnarray}
and $h_{\p} \equiv f_{\p}(1+f_{\p})$. The Coulomb logarithm, ${\mathcal L}$, is a divergent integral of the form ${\mathcal L}=\int_{q_{min}}^{q_{max}} dq/q$, where $q_{max}$ it typically of the order of the temperature, while $q_{min}$ is determined by screening effects, that is, $q_{min}\sim m_D$, where $m_D$ is the Debye screening mass.   We will take ${\mathcal L}$  to be a constant in our analysis, that is we ignore within the logarithm the time dependence of $m_D$ (we shall see that near the onset of BEC, $m_D$ is practically constant). A discussion of the effects of possible variations of ${\mathcal L}$ with time in a similar context can be found in \cite{Bjoraker:2000cf}.\footnote{Note a slight inconsistency in our treatment: we assume an implicit screening mass on the exchanged gluon, but ignore mass effects on the colliding ones. For massive gluons, both the matrix element and the tensorial structure in Eq.~(\ref{current0}) are modified. Also,  the criterion for condensation changes from  $\mu=0$ to  $\mu\approx m_D$, where $\mu$ is the chemical potential, which entails a modification of the dispersion relation at small momenta, from ultra-relativistic to non relativistic. However we do not expect these modifications to alter the main qualitative conclusions of this paper. }

The momentum integrations in Eq.~(\ref{current0}) can be performed and yield
\beq\label{current1}
{\mathcal J}(\p)=-2\pi^2\alpha^2\xi \,\left[ I_a \nab f(\p) + \frac{\p}{p}\, I_b\, f(\p)[1+f(\p)] \right].
\eeq
where  $\xi=18{\mathcal L}/\pi$, and $I_a$ and $I_b$ are the following integrals
\beq\label{definitionsI}
I_a\equiv \int\frac{d^3 p}{(2\pi)^3} f(\p)(1+f(\p)),\qquad I_b\equiv \int\frac{d^3 p}{(2\pi)^3}\frac{2f(\p)}{p}.
\eeq 
  The integral $I_a$ plays the role of a diffusion constant (to within numerical constants and the Coulomb logarithm, this is the proportionality coefficient between the current and the gradient of $f$), while  $I_b$ is proportional to the Debye screening mass, $m_D^2=2g^2N_cI_b$. Note that for an equilibrium distribution function $f_{eq}(p)$, the integrals $I_a$ and $I_b$ are simply related: $I_a=T\,I_b$. We have indeed
\beq\label{fequilibrium}
f_{eq}(p)=\frac{1}{{\rm e}^{(p-\mu)/T}-1}, \qquad  I_b[f_{eq}] =\!-\!\int\frac{d^3 p}{(2\pi)^3}\frac{\del f_{eq}}{\del p}=\frac{1}{T}\,I_a[f_{eq}],
\eeq
where we have performed an integration by part and used the property  $ {\del f_{eq}}/{\del p}=- f_{eq}(1+f_{eq})/T$.  It follows in particular that the current ${\mathcal J}$ vanishes for the Bose-Einstein distribution $f_{eq}$, and  that this distribution  is therefore a stationary solution of the Boltzmann equation.

At this point, we redefine the time and set 
\beq\label{scaletime1}
\tau\equiv 2\pi^2\alpha^2\xi t .
\eeq
 Then the transport equation reads simply
\beq\label{transporthat}
{\mathcal D}_\tau f(\tau,\p) =  \nab \cdot  \left[  I_a \nab f(\tau,\p) + \frac{\p}{p}\,  I_b\, f(\tau,\p)[1+f(\tau,\p)] \right].
\eeq
 Note that after the rescaling of the time, the transport equation contains no parameters. It is a universal equation, whose solutions are entirely determined by the initial conditions.  In the case where $f\ll 1$, one can replace $f(1+f)$ by $f$ in the equation, and similarly in the integral $I_a$. One then recovers the equation used earlier in Refs.~\cite{Mueller:1999pi,Bjoraker:2000cf} to study the thermalization of the quark-gluon plasma.
Finally, the non-local character of this equation is worth-emphasizing: although Eq.~(\ref{transporthat}) looks like a local partial differential equation for the function $f(\tau,\p)$, there is in fact a non-linear coupling with the entire solution through the integrals $ I_a$ and $ I_b$ which enter as coefficients of the equation.

In the case of an isotropic system, the transport equation takes the form
\beq\label{dfdtaufp}
{\del}_\tau f(\tau,p) = I_a\frac{\del^2 f}{\del p^2}+\frac{2 I_b}{p}\, f(1+f)+\left[ \frac{2 I_a}{p}+ I_b\,(1+2f)  \right]\frac{\del f}{\del p}.
\eeq
It can also be written in terms of the current ${\mathcal J}(\tau,p)$
\beq\label{dfdtauJ0}
\del_\tau f(\tau,p)=-\frac{1}{p^2}\frac{\del}{\del p}\left( p^2 {\mathcal J}(\tau,p)  \right),
\eeq
with ${\mathcal J}(\tau,p)$ given by
\beq\label{calJp0}
{\mathcal J}=-   I_a\, \partial_p f -  I_b\, f(1+f) .
\eeq

The equation (\ref{transporthat}) preserves the basic conservation laws of particle number, and energy:  it can indeed be easily verified through an explicit calculation that the particle density $n=\int_p f(p)$ and the energy density $\epsilon=\int_p pf(p)$ are independent of time. It is  instructive to perform this verification on the form (\ref{dfdtauJ0}) of the transport equation.  To do so, consider the number $n_{p_0}$ of particles (of a given species)  in a sphere of radius $p_0$. We have
\beq
n_{p_0}=\frac{1}{2\pi^2}\int_0^{p_0} dp\,p^2 f(\tau,p).
\eeq
By taking the time derivative of $n_{p_0}$, and using Eq.~(\ref{dfdtauJ0}), we obtain 
\beq
\frac{dn_{p_0}}{d\tau}=\frac{1}{(2\pi)^3}\left({\mathcal F}(0)-{\mathcal F}(p_0)\right),\qquad  {\mathcal F}(p)\equiv 4\pi p^2{\mathcal J}(p),
\eeq
where $ {\mathcal F}(p)$ is  the flux of particles crossing a sphere of radius $p$ (counted positively if the flow is away form the origin). Clearly ${\mathcal F}(p_0)$ vanishes as $p_0\to \infty$ and, in the absence of condensation, ${\mathcal F}(0)=0$. This ensures that particle number is conserved.  
Similarly,  for the energy density $\epsilon_{p_0}$ contained in a sphere  of radius $p_0$,
\beq 
\epsilon_{p_0}=\frac{1}{2\pi^2}\int_0^{p_0} dp\, p^3f(p),
\eeq
we have
\beq
(2\pi)^3 \frac{d\epsilon_{p_0}}{d\tau}= \left.p{\mathcal F}(p)\right|_{p=0}-p_0 {\mathcal F}(p_0)+4\pi \int_0^{p_0} dp \,p^2{\mathcal J}(p).
\eeq
The arguments used above for the density can be invoked here to show that the first two terms vanish as $p_0\to\infty$. 
As for the integral, it  can be written as
\beq
-\frac{1}{4\pi}\int_0^{p_0} dp \,{\mathcal F}(p)=I_a \int_0^{p_0}dp \,p^2\, \del_p f(p)+I_b\int_0^{p_0} dp\, p^2 f(1+f).
\eeq
Then, recalling the definitions (\ref{definitionsI}) of $I_a$ and $I_b$, we get an exact cancellation in the r.h.s. as $p_0\to\infty$. This is because  $\int_0^{p_0} dp\, p^2 \del_p f=-\int_0^{p_0} dp\, p^2 (2f/p)$ to within surface terms that vanish in the limit $p_0\to\infty$.  It is worth noticing that while the particle number conservation follows just from the fact that the time derivative of $f(p)$ is the divergence of a current, the conservation of energy involves explicitly the integrals $I_a$ and $I_b$. 

\section{Initial conditions and two types of solutions}

We shall be interested in this paper in following the time evolution of the distribution function $f(\tau,\p)$, as given by the transport equation (\ref{dfdtaufp}), starting from some initial  condition. We shall assume that the phase space density is independent of spatial coordinates and is isotropic in momentum.

For the typical initial conditions that will be considered in this paper, one expects generally  two types of solutions: some lead at late time to a thermal Bose-Einstein distribution function; others show a transition, \emph{in a finite time}, to a Bose-Einstein condensate. Which solution is attained depends typically on one parameter characterizing the considered family of initial conditions.   To illustrate this point, consider the static, isotropic case, with a particular family of initial conditions of the generic form $f(p)=f_0\, g(p/Q_s) $, where $Q_s$ fixes the scale of momenta, and $g(q)$ is normalized so that $\int_0^\infty {\rm d}q\, q^2\, g(q)=1/3$. A particularly simple distribution is that inspired by the color glass (CGC)  picture \cite{Blaizot:2011xf}
\begin{eqnarray} \label{eq_glasma_f}
f( p ) = f_0 \, \theta(1 - p/Q_s ),
\end{eqnarray}
with $Q_s$ the saturation scale.  With this initial distribution, we have 
\beq\label{initialparam}
 \epsilon_0  = f_0\, \frac{Q_s^4}{  8\pi^2},\qquad n_0=f_0\, \frac{Q_s^3}{6\pi^2},\qquad
n_0 \, \epsilon_0^{-3/4} = f_0^{1/4} \, \frac{2^{5/4}}{3\, \pi^{1/2}} ,
\eeq
with $\epsilon_0$ and $n_0$ the initial energy density and number density, respectively. 
The value of the parameter $n \, \epsilon^{-3/4}$ that corresponds to the onset of Bose-Einstein condensation, i.e., to an equilibrium state with vanishing chemical potential,  is obtained by taking for $f(p)$ the  ideal  distribution for massless particles at temperature $T$. One gets then $\epsilon_{SB}=(\pi^2/30)\, T^4$ and $n_{SB}=(\zeta(3)/\pi^2)\, T^3$, so that 
\begin{eqnarray}\label{overpopparam}
n \, \epsilon^{-3/4}  |_{SB} = \frac{30^{3/4}\, \zeta(3)}{\pi^{7/2}} \approx 0.28.
\end{eqnarray}
Comparing with $n_0 \, \epsilon_0^{-3/4} $ in Eq.~(\ref{initialparam}), one  sees that when $f_0$ exceeds the value $f_0^c \approx 0.154$, the initial distribution (\ref{eq_glasma_f}) contains too many gluons to be accommodated in an equilibrium Bose-Einstein distribution: the equilibrium state will  contain a  Bose-Einstein condensate.\footnote{ 
Note that we have assumed in this estimate a single species of gluon. In principle we should multiply both $\epsilon$ and $n$ by the degeneracy factor $\nu=16$. This will change the over-occupation parameter by a factor $\nu^{1/4}$. However the same factor would also enter the ideal gas values, so that $f_0^c$ is unchanged.} The value of $f_0^c$ depends of course on the function $g$. For instance, for $g(p/Q_s)\propto\delta(1-p/Q_s)$,  $f_0^c\approx 0.36$. Note that over-occupation does not require necessarily large values of $f_0$, in fact the values just quoted are smaller than unity. It follows therefore that the situation of over-occupation will be met for generic values of $\alpha_s$. For instance, for $\alpha_s\simeq 0.3$, $f_0=1/\alpha_s$ is significantly larger than $f_0^c$ for a wide class of initial conditions.

When $f_0<f_0^c$ the system is initially under-populated, and will evolve to a Bose-Einstein distribution function (see Eq. ~(\ref{fequilibrium})),  with a finite (negative) chemical potential. This evolution, that takes place at fixed number density and energy density, is accompanied by a growth of  the entropy 
\beq
{\mathcal S}=\int \frac{d^3\p}{(2\pi)^3}\left[ (1+f)\ln (1+f)-f\ln f   \right].
\eeq
A simple calculation yields 
\beq\label{entropyrate}
\frac{d{\mathcal S}}{d\tau}=\int \frac{d^3\p}{(2\pi)^3} \,\frac{\del f}{\del \tau}\,\ln\frac{1+f}{f}=-\int  \frac{d^3\p}{(2\pi)^3} \,{\mathcal J}(p)\,\frac{1}{f(1+f)}\frac{\del f}{\del p},
\eeq
which shows that the entropy ceases indeed to grow when $f $ is an equilibrium distribution function: this follows immediately from the conservation laws that we have just discussed, or alternatively, from the vanishing of the current ${\mathcal J}(p)$  in equilibrium. In the present context, the study of the current ${\mathcal J}(p)$ in momentum space turns out to be a useful tool to follow the process of  thermalization.

The current ${\mathcal J}(p)$ is given in Eq.~(\ref{calJp0}). It is composed of two terms. The first term, proportional to $ I_a$ and to the gradient of the distribution can be associated with a diffusion process that takes particles from region of high phase-space density to region of low phase-space density. The other component of the current, proportional to $ I_b$ is simply proportional to the magnitude of $f$ (more precisely to $f(1+f)$). It is negative, and drives therefore particles towards the low momentum region, and it is most intense in the regions where $f$ is large. Initially, the diffusion current will concern only the high momentum modes (for the CGC initial conditions). However, when low momentum modes are substantially populated, a competition sets in between the two components of the current, leading eventually to equilibrium. 

In order to get some more orientation on how the system evolves towards equilibrium,  it is instructive to evaluate the integrals $I_a$ and $I_b$ at the initial time. We get
\beq
 I_a^0=\frac{Q_s^3}{6\pi^2}\,f_0(1+f_0),\qquad  I_b^0=\frac{Q_s^2}{2\pi^2}\,f_0.     
\eeq
It is also useful to introduce the following quantities
\beq\label{T*0}
T^*\equiv\frac{I_a}{I_b}=\frac{1+f_0}{3}\,Q_s,\qquad  \alpha^*=\ln \left( 1+\frac{1}{f(0)}   \right).
\eeq
\begin{figure*}[h!]
	\begin{center}
		\includegraphics[width=6cm]{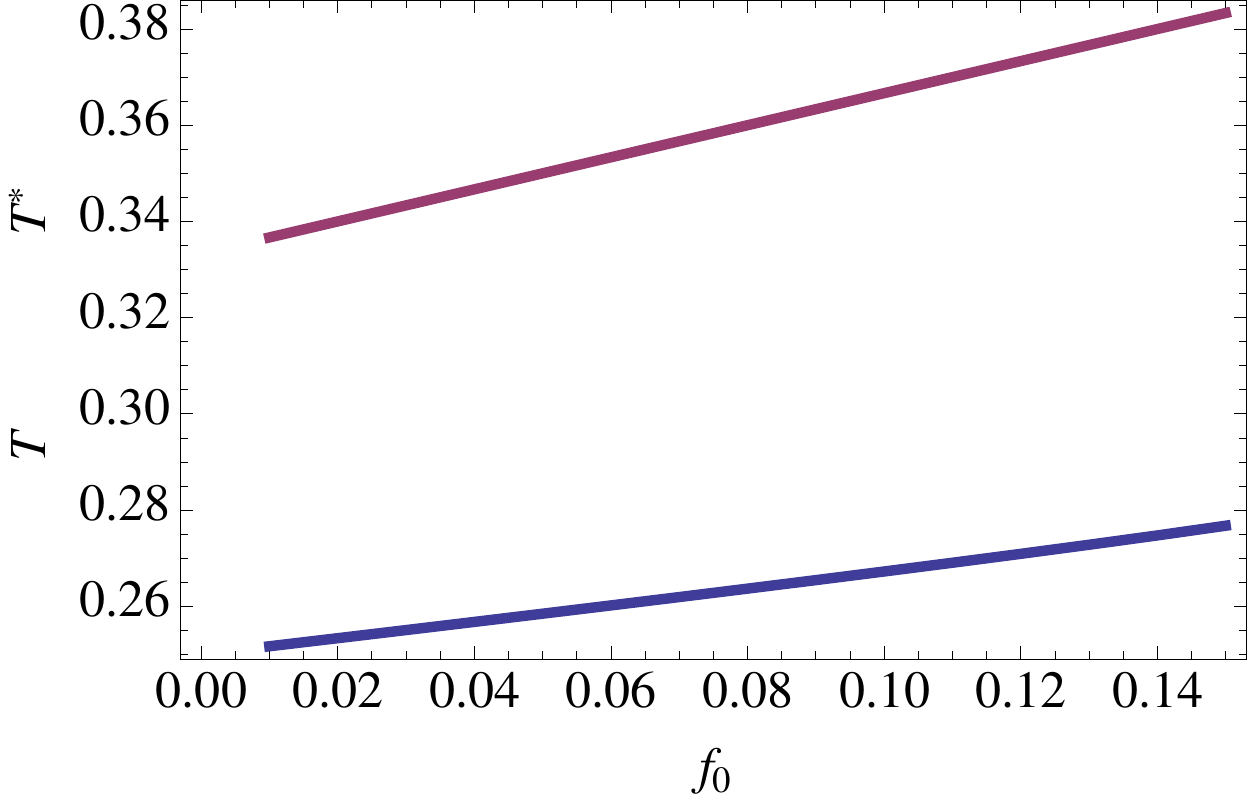} \hspace{0.2cm}
	\includegraphics[width=6cm]{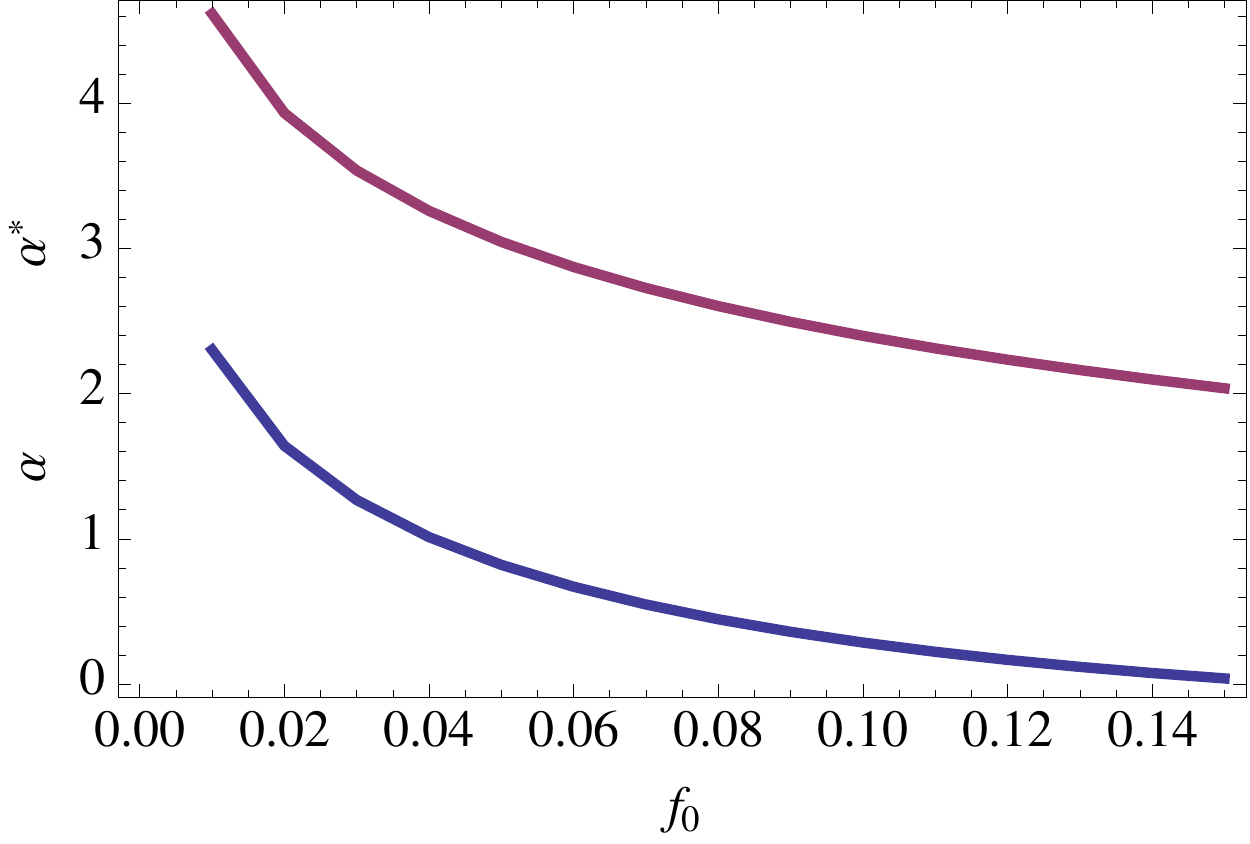}
		\caption{(color online) Left: the equilibrium ($T$) and effective  ($T^*$) temperatures as a function of $f_0$.  Right: $\alpha$ and $\alpha^*$ as a function of $f_0$. Note that $\alpha$ vanishes at the critical value $f_0^c=0.154$. }
		\label{fig:Tandmueq}
	\end{center}
%	\vspace{-1cm}
\end{figure*}

The quantity $T^*$ is an effective temperature. This interpretation will be made more transparent in the next section, but the fact that it goes over  to the equilibrium temperature as the system thermalizes is clear from Eq.~(\ref{fequilibrium}) and the discussion around.  Similarly, $\alpha^*$ will be seen to be related to a (negative) effective chemical potential $\mu^*$,  $\alpha^*=|\mu^*|/T^*$, and is a measure of the value of the distribution function at zero momentum. For the initial distribution (\ref{eq_glasma_f}), $\alpha^*$ is given by Eq.~(\ref{T*0}) with $f(0)=f_0$. For the equilibrium distribution (see Eq.~(\ref{fequilibrium})),   $f_{eq}(0)=1/(\exp(-\mu/T)-1)$,  so that,  in equilibrium, $\alpha=|\mu|/T$.
In Fig.~\ref{fig:Tandmueq}, we show the variations with $f_0$ of the equilibrium values $T$ and $\alpha$. One sees that $T$ and $T^*$ follow the same smooth variations with $f_0$, and the same holds for $\alpha$ and $\alpha^*$, with $\alpha$ vanishing at the critical value of $f_0$. For a given initial $f_0<f_0^c$, one expects therefore thermalization to be generically accompanied by a decrease of both $T^*$ and $\alpha^*$.

When $f_0>f_0^c$,  the system is initially over-populated and the equilibrium state contains a  Bose-Einstein condensate. That is, the equilibrium distribution is of the form
\beq\label{fcondensate}
f_{eq}(p)=\frac{1}{{\rm e}^{p/T}-1}+n_c\,\delta^{(3)}(\p),\qquad n=\frac{\zeta(3)}{\pi^2}\, T^3+n_c, 
\eeq
with $n_c$  and $T$ determined by  the  initial density and energy density.  
We can calculate the equilibrium  temperature from the initial energy density
\beq
T=Q_s\left( \frac{15}{4\pi^4}   \right)^{1/4}\,f_0^{1/4},
\eeq
while $\alpha=\mu=0$. The value of the condensate can be extracted from the expression of the density given in Eq.~(\ref{fcondensate}) above. The variation of $T$ with $f_0$  is displayed in Fig.~\ref{fig:Tover}. One sees that the presence of the condensate makes it  milder than in the under-populated case, while $T^*$ remains linear in $f_0$. It follows that the gap between the initial effective temperature and the equilibrium one increases. Note however that the equilibrium temperature is not the temperature at which Bose condensation will set in. As we shall see,  the temperature at the onset for BEC is actually higher than $T$ (and not too much lower than $T^*$ when $f_0$ is  large). 
\begin{figure*}[h]
	\begin{center}
		\includegraphics[width=6cm]{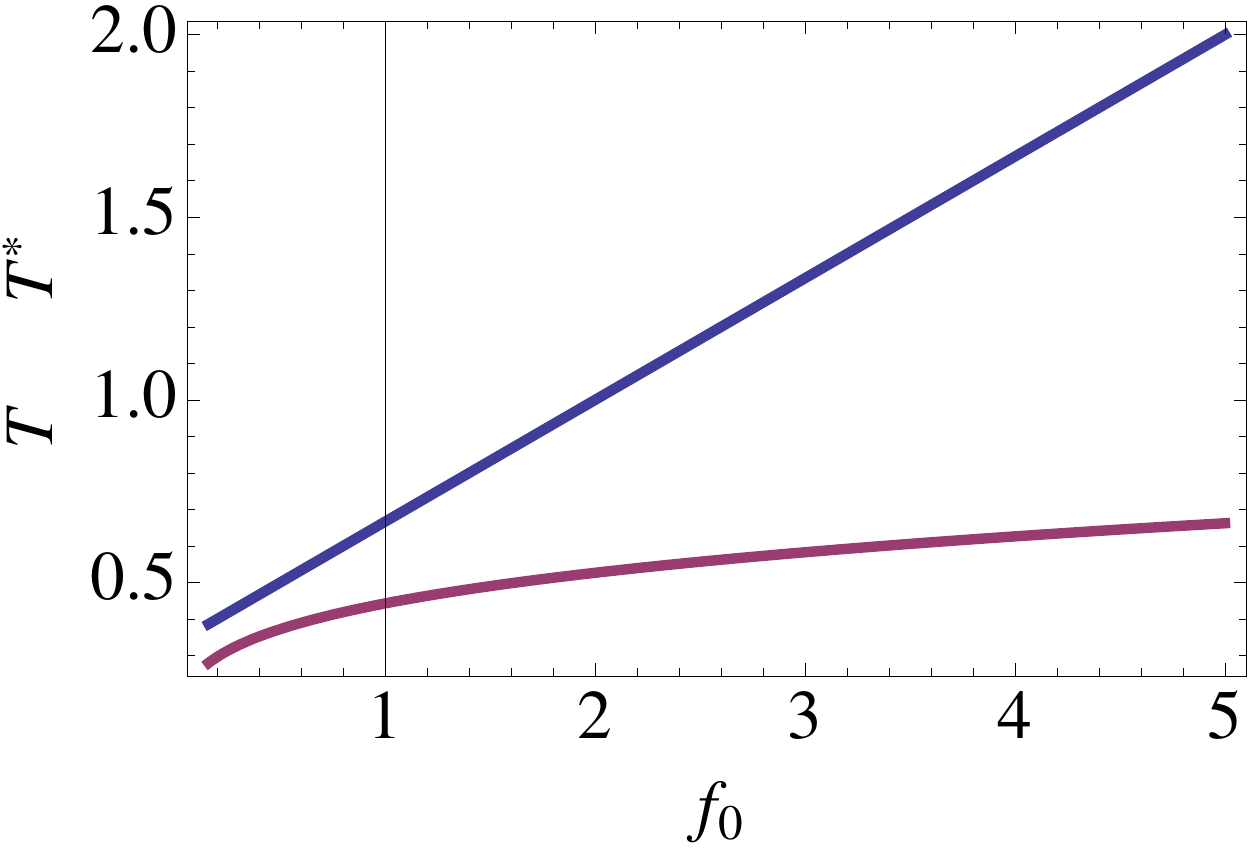} \hspace{0.2cm}
		\caption{(color online) The equilibrium temperature (lower curve) and the effective temperature (upper curve) as a function of $f_0$, for the over-populated case, $f_0>f_0^c\approx 0.154$. }
		\label{fig:Tover}
	\end{center}
%	\vspace{-1cm}
\end{figure*}

In the present study, we are not in a position to follow the evolution of the system beyond the condensation point. This would require modifying the transport equation to take into account the time evolution of the condensate and the coupling between the condensate and the non condensed particles.  Such a study will be presented in a forthcoming publication. Here, we shall only study the onset of condensation and provide numerical evidence that the over-occupied system reaches this  point  in a finite time.

%%%%%%%%%%%%%%%%%%%%%%%%%%%%%%%%
\section{The under-occupied case: approach to thermal equilibrium}
%%%%%%%%%%%%%%%%%%%%%%%%%%%%%

In this section, and the next one, we present a detailed numerical study of the solution of the transport equation (\ref{dfdtauJ0}) in the two regimes $f_0<f_{0}^c$ (this section) and $f_0>f_{0}^c$ (next section). 
We  consider  CGC motivated initial distributions of the form given by Eq~(\ref{eq_glasma_f}). However, in order to avoid the divergence in $\partial_p f(p)$ that would occur with the step-function distribution,  we use  in our numerical studies the following smooth function
\begin{eqnarray}\label{finitCGC}
f(\tau_0,p) = f_0 \, \left[ \theta(1 - p/Q_s ) + \theta(p/Q_s-1) e^{-a\, (p/Q_s-1)^2} \right],
\end{eqnarray}
with the numerical constant $a=10$. For this particular family of initial distributions,  the critical value of $f_0$ is $f_0^c=0.1675$.  We use a momentum grid with typically  $0\le p/Q_s \le 5$, and a grid size $\Delta p \le 0.01$ (finer meshes have also been used in some cases). The time step is chosen of the order of $10^{-5}$. The numerical computation perfectly conserves particle number; it also conserves energy to better than $0.05\%$  in the longest evolution. 

In order to facilitate the comparison of different initial conditions, it is convenient to redefine the time $\tau$ so as to incorporate the change in time scale caused by changes of $f_0$, i.e., $\tau\rightarrow \tau\,f_0(1+f_0)$. Thus, in all results presented in this section and the next, 
\beq\label{taunew}
\tau=2\pi^2\alpha^2\xi\,f_0(1+f_0)\, t.
\eeq
This redefinition of $\tau$ makes it explicit  that in the regime where $f_0\sim 1/\alpha\gg 1$, the coupling constant drops out from the relation between $t$ and $\tau$.
We absorb the change in the normalization of $\tau$ by simultaneously redefining the integrals $I_{a,b}$: we  set $\hat I_{a,b}=I_{a,b}/f_0(1+f_0)$.
The equations  (\ref{dfdtauJ0}) and (\ref{calJp0}) become then
\beq\label{dfdtauJ}
\del_\tau f=-\frac{1}{p^2}\frac{\del}{\del p}\left( p^2 {\mathcal J}(p)  \right),\quad {\mathcal J}(p)=-  \hat I_a\, \partial_p f(p) - \hat I_b\, f(p)[1+f(p)] .
\eeq
Figure \ref{fig:distr_f01} shows the typical time evolution of the distribution function, starting from an under-populated initial condition (here, with $f_0=0.1$).  The approach to equilibrium proceeds by a gradual modification of the momentum distribution, the momentum region around $Q_s$ being gradually depleted in favor of an increase of the population of softer and harder modes. A striking  feature, particularly visible in the linear plot (left panel of Fig.~\ref{fig:distr_f01}),   is the rapid growth of the population of low momentum modes. This is analyzed in the next subsection.

\begin{figure*}[h]
	\begin{center}
		\includegraphics[width=6cm]{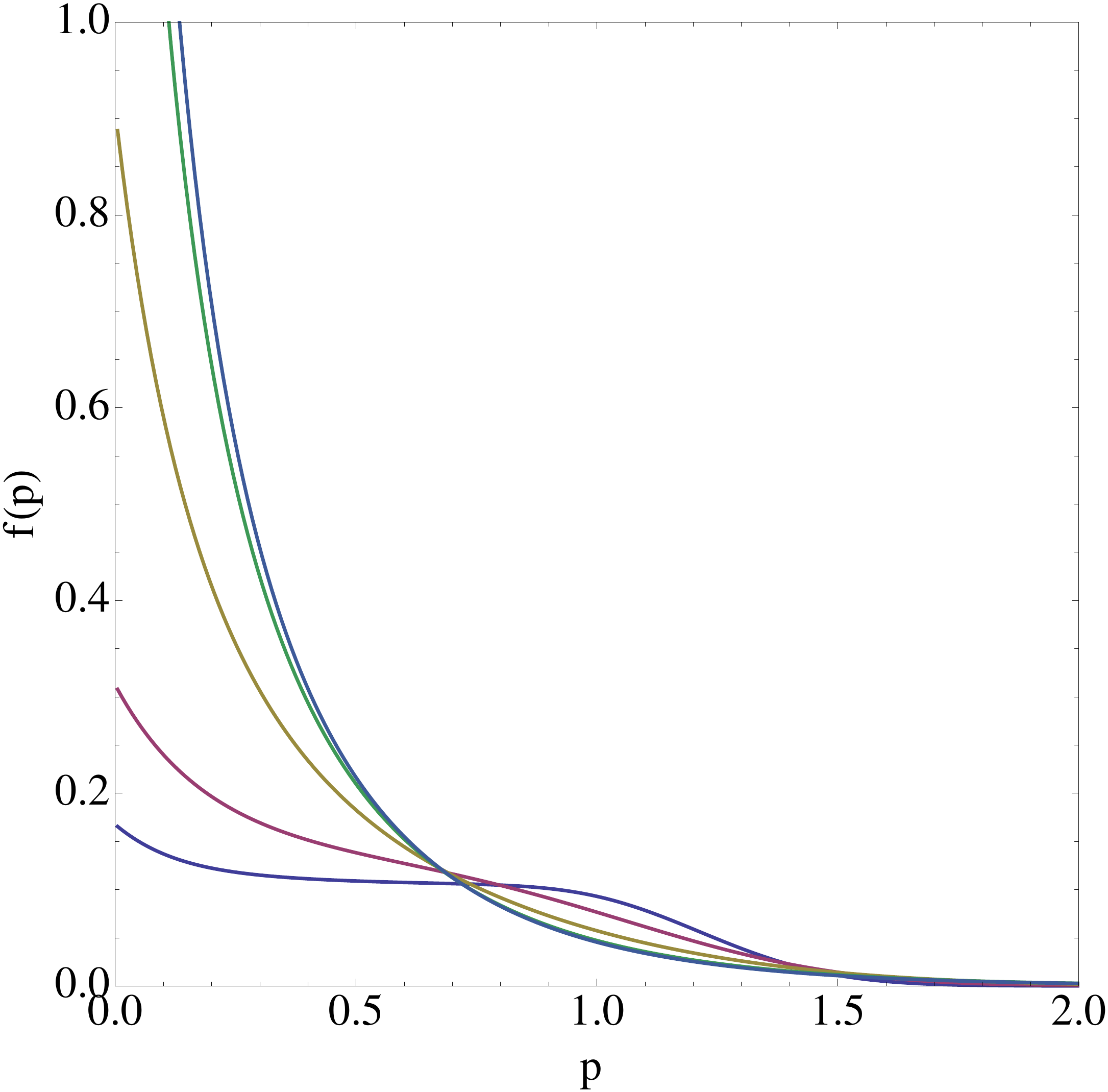} \hspace{0.2cm}
		\includegraphics[width=6cm]{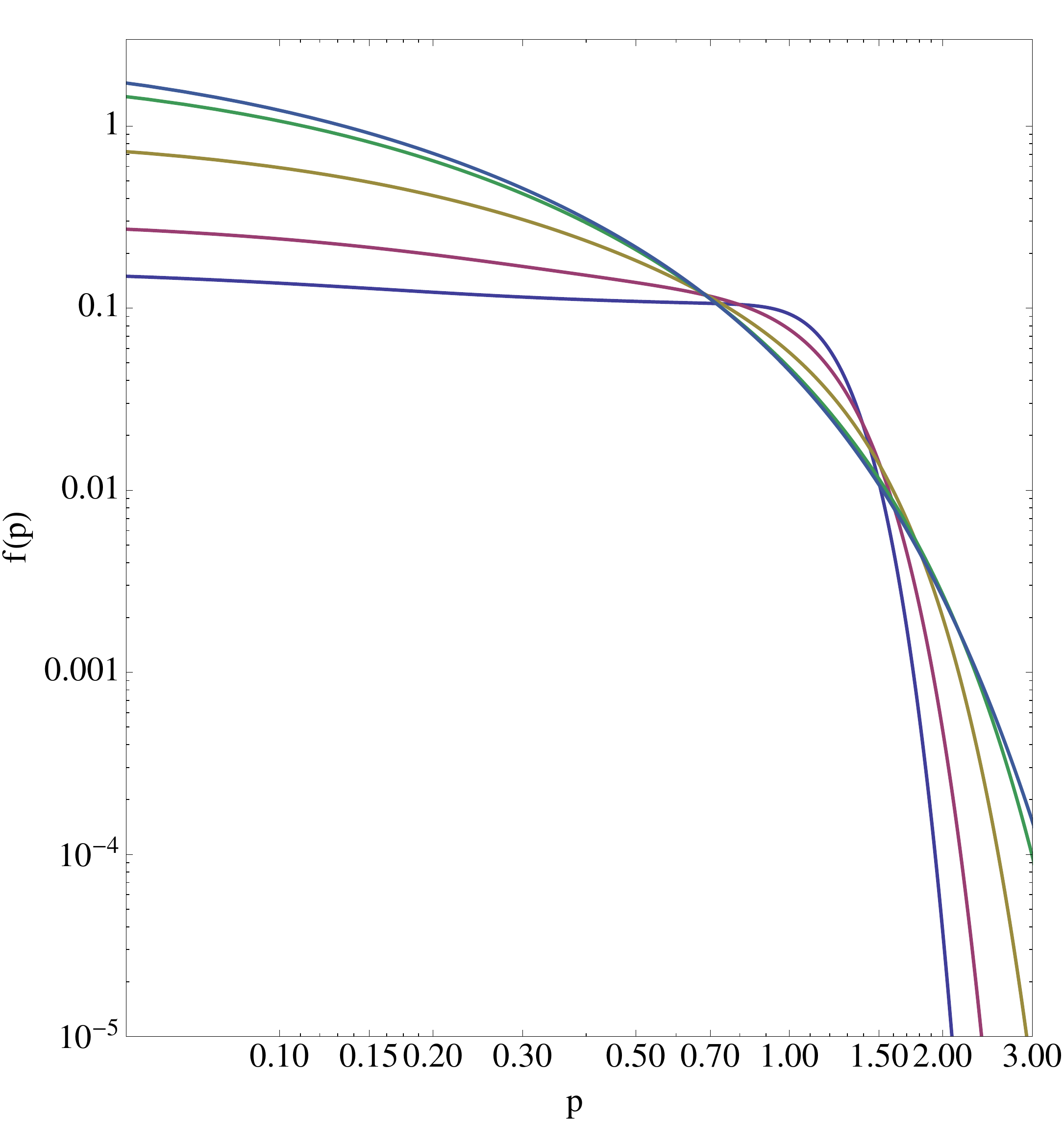}
		\caption{The  distribution function $f(\tau,\p)$ as a function of $p$, for various times, from an early time till the time where thermalization is nearly completed.  Left: linear plot. Right: double logarithmic plot. [$f_0=0.1$]}
		\label{fig:distr_f01}
	\end{center}
%	\vspace{-1cm}
\end{figure*}

%%%%%%
\subsubsection{The growth of the  small momentum modes}
%%%%%%

We have already emphasized in the previous section that the current contains two contributions, one that can be associated to diffusion and that is proportional to the gradient of the distribution, the other being simply proportional to $f(1+f)$, and negative. A plot of the initial current is shown in Fig.~\ref{fig:fz01_current}, where the large and negative component is clearly visible. The (positive) diffusion part  only  affects the high  ($p\gtrsim Q_s$) momentum modes where $f$ presents a strong (negative) gradient. This form of the initial current is responsible for the rapid growth of the population of the small momentum modes.

 At later times,  the form of the current is strongly constrained by the transport equation. To  see that, let us integrate Eq.~(\ref{dfdtauJ}) from 0 to $p_0\ll Q_s$. We  get
\beq\label{smallpeqn}
\del_\tau \int_0^{p_0} dp\,p^2f(p)=-p_0^2\, {\mathcal J}(p_0),
\eeq
where we have assumed that $p^2{\mathcal J}(p)\to 0$ as $p\to 0$ (in the absence of condensation, there is no net flux of particles at the origin). This equation implies that as long as $f(p)$ and its time derivative  are regular as $p\to 0$, then ${\mathcal J}(p) $ vanishes linearly with $p$. This is because, for $p_0$ small enough, we may assume $f$ to be constant $\approx f(0)$ inside the sphere of radius $p_0$ and integrate  the l.h.s. of Eq.~(\ref{smallpeqn}) to get 
\beq\label{currentfdot}
\frac{\del f(0)}{\del \tau}=-\frac{3}{p_0}{\mathcal J}(p_0).
\eeq
This condition is non trivial. It implies a subtle interplay, already alluded to,  between the global information contained in the integrals $I_a$ and $I_b$, and the local properties of the solution. 
\begin{figure*}[t]
	\begin{center} 
	\includegraphics[width=6cm]{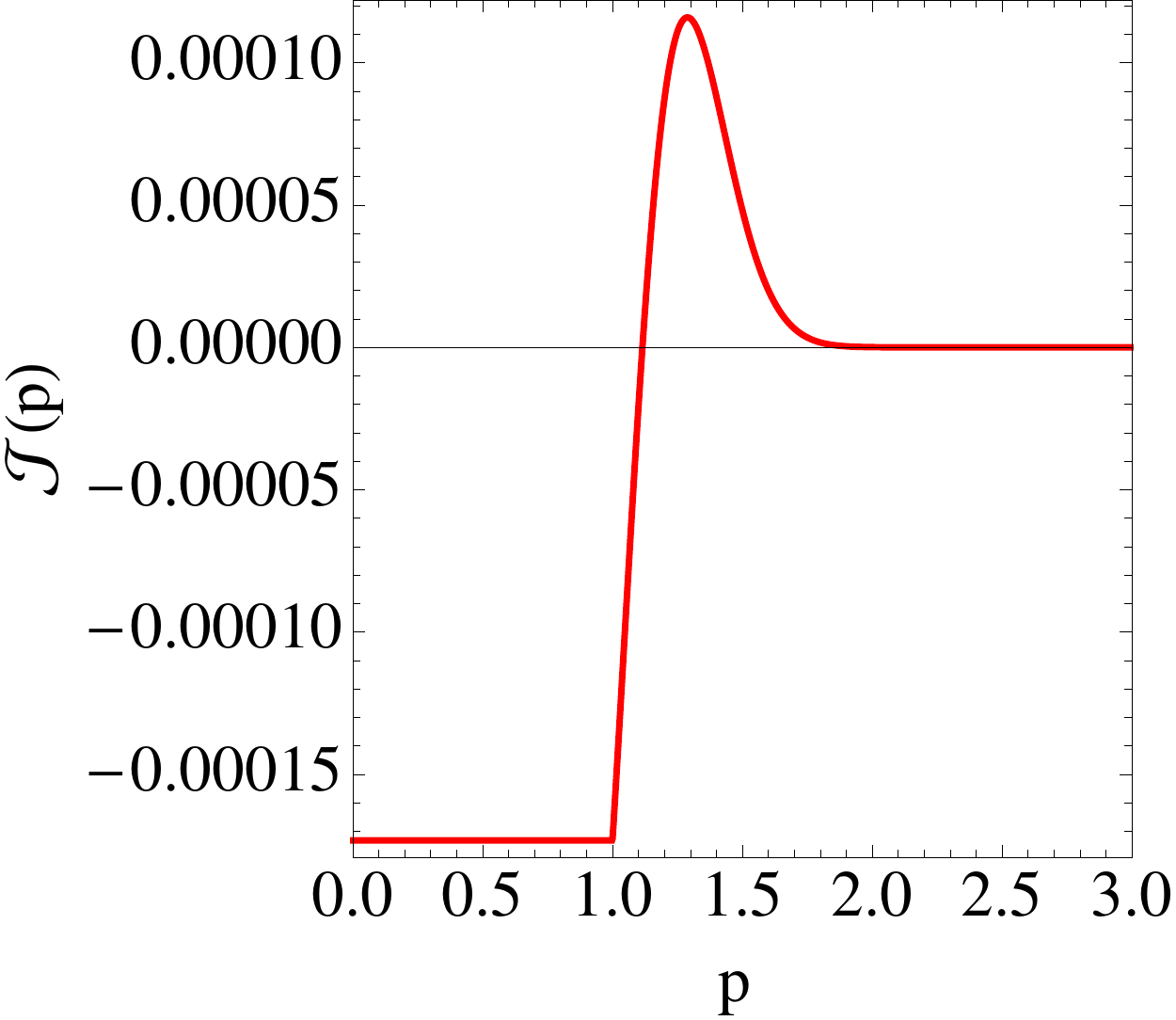}\hspace{0.2cm}
	\includegraphics[width=6cm]{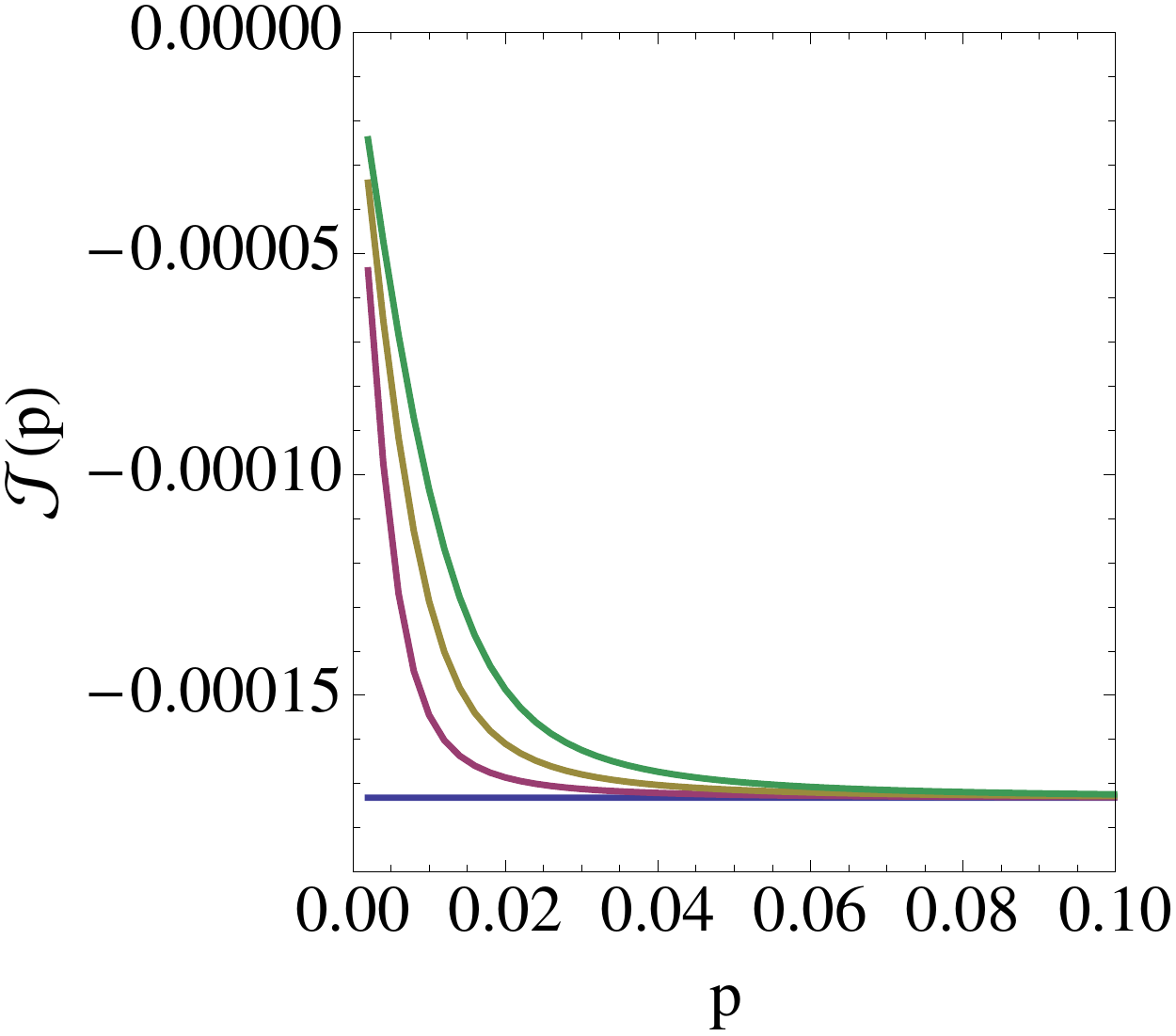}
		\caption{(color online) Left: the  current ${\mathcal J}(p)$ at initial time. Right: the current ${\mathcal J}(p)$ for small $p$, at  very early times, from $\tau=0$ to $\tau\simeq 0.0002$.  ($f_0=0.15$).  }
		\label{fig:fz01_current}
	\end{center}
%	\vspace{-1cm}
\end{figure*}

To understand  better how this condition (\ref{currentfdot}) can be satisfied by the current (\ref{calJp0}), we observe that at small momenta, the distribution function quickly adjusts  its  shape  to a local distribution function. This is because the collision rate is large at small momenta: in fact,  the collision rate $(1/f)\del_\tau f$ diverges as $1/p$ at small $p$ when ${\mathcal J}$ is constant.\footnote{The collision rate is also large at large $p$, but then the distribution function is very small. Besides, for large momenta, small angle collisions  do not change momenta by large amounts, so that they are less efficient to drive the high momentum particles to equilibrium. } Furthermore, collisions change small momenta by amounts of order one, hence can drive the soft modes quickly to equilibrium. Note finally that soft particles scatter predominantly on hard ones (because these are most numerous), rather than on soft ones. Thus we may expect the small angle approximation to remain reasonably accurate (it would not be so if collisions among soft particles were dominant). Let us then assume that at small $p$, $f(p)$ deviates slightly form an equilibrium distribution, for which the current vanishes.  That is, we assume that for small $p$, $f(p)$ is of the form
\beq
f(p)=f^*(p)+g(p), \qquad g(0)=0,
\eeq
where $f^*(p)$ is chosen so that 
\beq\label{localeqforf}
0=\hat I_a\frac{\del f^*}{\del p}+\hat I_b f^* (1+f^*).
\eeq
At this point, we assume that one can ignore the feedback of the change of the distribution function on the values of the integrals $\hat I_a$ and $\hat I_b$.\footnote{We are making here an approximation akin to an adiabatic approximation:  the rate of change of the distribution function is much larger than that of the integrals $\hat I_a$ and $\hat I_b$, which play the role of slow variables in this problem.} In other words, we regard Eq.~(\ref{localeqforf}) as a simple differential equation for $f^*$, at fixed $\hat I_a$ and $\hat I_b$. This equation  implies  that $f^*$ is a Bose-Einstein equilibrium distribution function
\beq
f^*(p)=\frac{1}{{\rm e}^{(p-\mu^*)/T^*}-1}, 
\eeq
with $T^*$ the effective temperature already introduced in the previous section, namely
\beq
T^*\equiv \frac{I_a}{I_b},
\eeq
while the value of the effective chemical potential  $\mu^*$ follows from the condition $g(0)=0$, i.e., 
\beq
f(0)=f^*(0)=\frac{1}{{\rm e}^{-\mu^*/T^*}-1},\qquad  -\frac{\mu^*}{T^*}=\ln\left(1+\frac{1}{f(0)}\right)=\alpha^*.
\eeq
We emphasize that the auxiliary equilibrium distribution function  $f^*$ is not meant to represent the entire distribution, but is just  an approximation to the distribution function at small momenta. In particular, the parameters $T^*$ and $\alpha^*$ are not determined by the energy density and the number of particles, but simply from local properties of the solution at small momenta. In fact, since this is to be used only at very small momenta, we may replace $f^*$ by its `classical' approximation, $f^*(p)\approx T^*/(p-\mu^*)$.

By using the explicit expression of the current in terms of $\hat I_a$ and $\hat I_b$, Eq.~(\ref{dfdtauJ}), we get from Eq.~(\ref{currentfdot})
\beq\label{currentcorrection}
p_0\frac{\dot f(0)}{3}=\hat I_a\frac{\del f}{\del p}+\hat I_b f(1+f)=\hat  I_a\frac{\del g}{\del p}+\hat I_b\left(  g(1+2 f^*)+g^2 \right),
\eeq
where all functions are to be evaluated at $p=p_0$. Since $g(p)$ is regular at small $p$ and vanishes at $p=0$, one can easily solve this equation by iteration, with the leading term obtained by keeping only the contribution of $\del g/\del p$ in the expression of the current. One then gets trivially $g(p)=p^2 \dot f(0)/(6\hat I_a)$,
that is 
\beq
f(p_0)\simeq f^*(p_0)+\frac{p_0^2\dot f(0)}{6I_a}\approx \frac{T^*}{p_0+|\mu^*|}+\frac{p_0^2\dot f(0)}{6\hat I_a}.
\eeq
This equation provides an accurate description of the small $p$ behavior of the distribution function. It  is used in the fitting procedure that determines $\mu^*$ (numerically we obtain thus a better determination of $\mu^*$ than by using simply the relation $f(0)=f^*(0)\approx T^*/|\mu^*|$). Note however that the $p^2$ correction to the distribution function, while essential to satisfy Eq.~(\ref{currentcorrection}),  does not contribute to the left hand side of Eq.~(\ref{smallpeqn}). We shall return to this discussion later when we analyze what happens when $\mu^*\to 0$, that is, when one approaches the condensation point.

At this stage, we note that the property of the current to be linear at small $p$ cannot be satisfied at very early time. Initially, indeed, the current is negative and nearly constant at small $p$. In fact we see from Eq.~(\ref{dfdtauJ}) that a current constant at small $p$ entails the development of a $1/p$ contribution to  $f(p)$, which when acted upon by the Laplacian present in the kinetic equation, generates a $\delta$-function singularity. However, the singularity is not strong enough to allow for a condensate to develop. What happens instead is a transient regime where gradually the current goes into its expected linear behavior. This transient regime is illustrated in the right panel of  Fig.~\ref{fig:fz01_current}.

\begin{figure*}[t]
	\begin{center}
		\includegraphics[width=6cm]{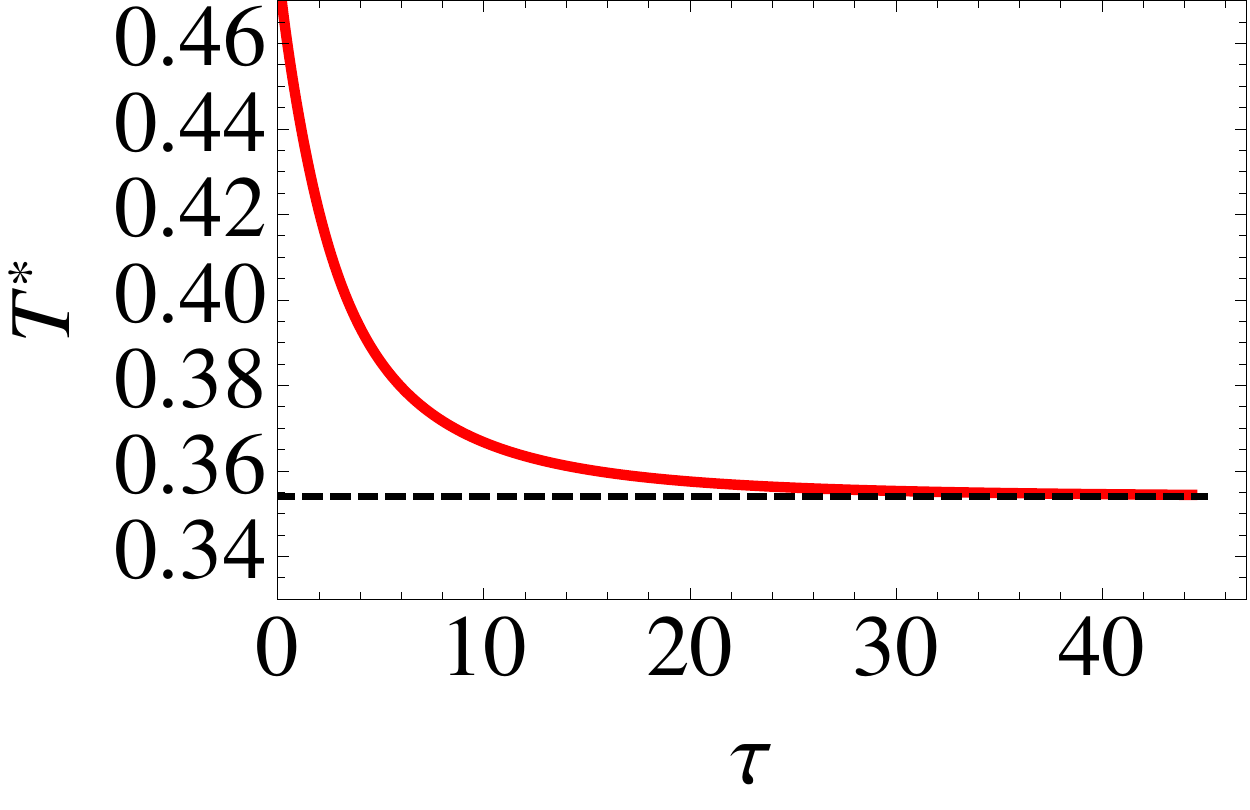} \hspace{0.2cm}
		\includegraphics[width=6cm]{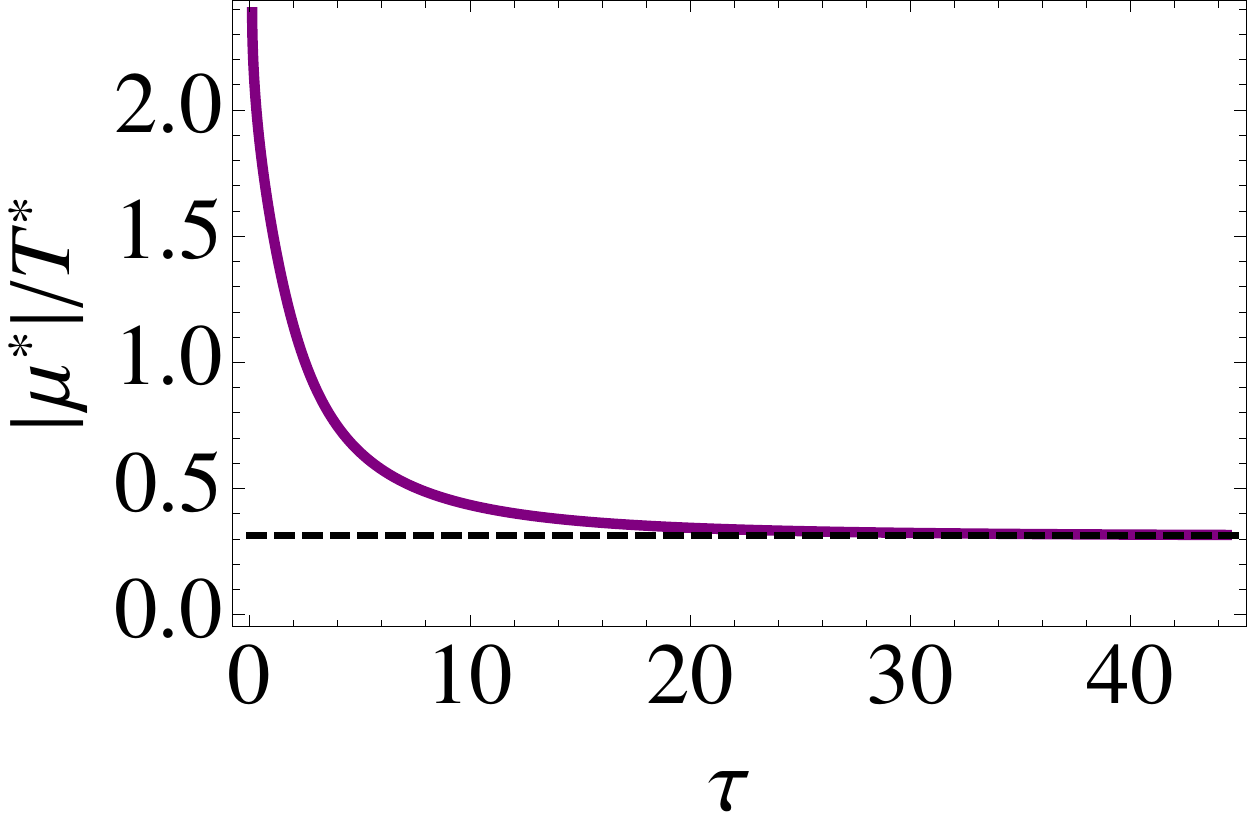}
		\caption{(color online)  The effective temperature $T^*$, and the quantity $\alpha^*=|\mu^*|/T^*$  as a function of $\tau$. Under-populated case, with $f_0=0.1$:  the equilibrium temperature is  $T=0.354$ and the equilibrium $\alpha$ is $\alpha=0.315$. These equilibrium values are indicated by the horizontal dashed lines.}
		\label{fig_alpha*}
	\end{center}
	%\vspace{1cm}
\end{figure*}

\subsubsection{Thermalization in the under-populated regime}

We turn now to the description of the overall evolution towards equilibrium in the under-populated case  $f_0<f_0^c$.  Fig.~\ref{fig_alpha*} displays the variations of the effective temperature $T^*$, and the effective chemical potential as the quantity $\alpha^*=|\mu^*|/T^*$. As discussed  at the end of the previous section, these two quantities must decrease in order for the system to reach equilibrium. As seen from Fig.~\ref{fig_alpha*} this decrease is regular, and the time variations of $T^*$ and $\alpha^*$ provide indications for when thermalization is approximately achieved: thus, both $T^*$ and $\alpha^*$ have almost reached their equilibrium values, $T=0.354$ and $\alpha=0.315$ at $\tau\gtrsim 20$. Note however  that this concerns only the lowest moments of the distribution function. The high momentum tail of the distribution reaches its equilibrium value only asymptotically. For instance, we have verified that the higher moments of the distribution evolve more slowly than $T^*$ and $\mu^*$.

It is also interesting to consider the  current  ${\mathcal J}(p)$ as well as the flux ${\mathcal F}(p)=4\pi p^2 {\mathcal J}(p)$ as a function of time. These quantities are displayed in Fig.~\ref{fig:fz01_flux}. The plot of the flux (left panel of Fig.~\ref{fig:fz01_flux}) provides a clear illustration of how particles move in momentum space as a function of time: from the initial momentum $Q_s$ towards small and large momenta. As time passes, the flux decreases regularly and becomes vanishingly small by the time $\tau\approx 20$, i.e., by the time at which $T^*$ and $\mu^*$ have relaxed to the their equilibrium values (see Fig.~\ref{fig_alpha*}). 
A similar behavior is observed for the current (right panel of Fig.~\ref{fig:fz01_flux}), with a linear slope almost constant at small momenta: this, according to Eq.~(\ref{currentfdot}),  indicates that $f(0)$ grows almost linearly with time, until late times where the slope eventually vanishes as the system relaxes to equilibrium. This picture continues to hold until $f_0$ reaches its critical value $f_0^c$, without any noticeable changes in time scales (measured with $\tau$ defined in Eq.~(\ref{taunew})).

\begin{figure*}[t]
	\begin{center}
		\includegraphics[width=6cm]{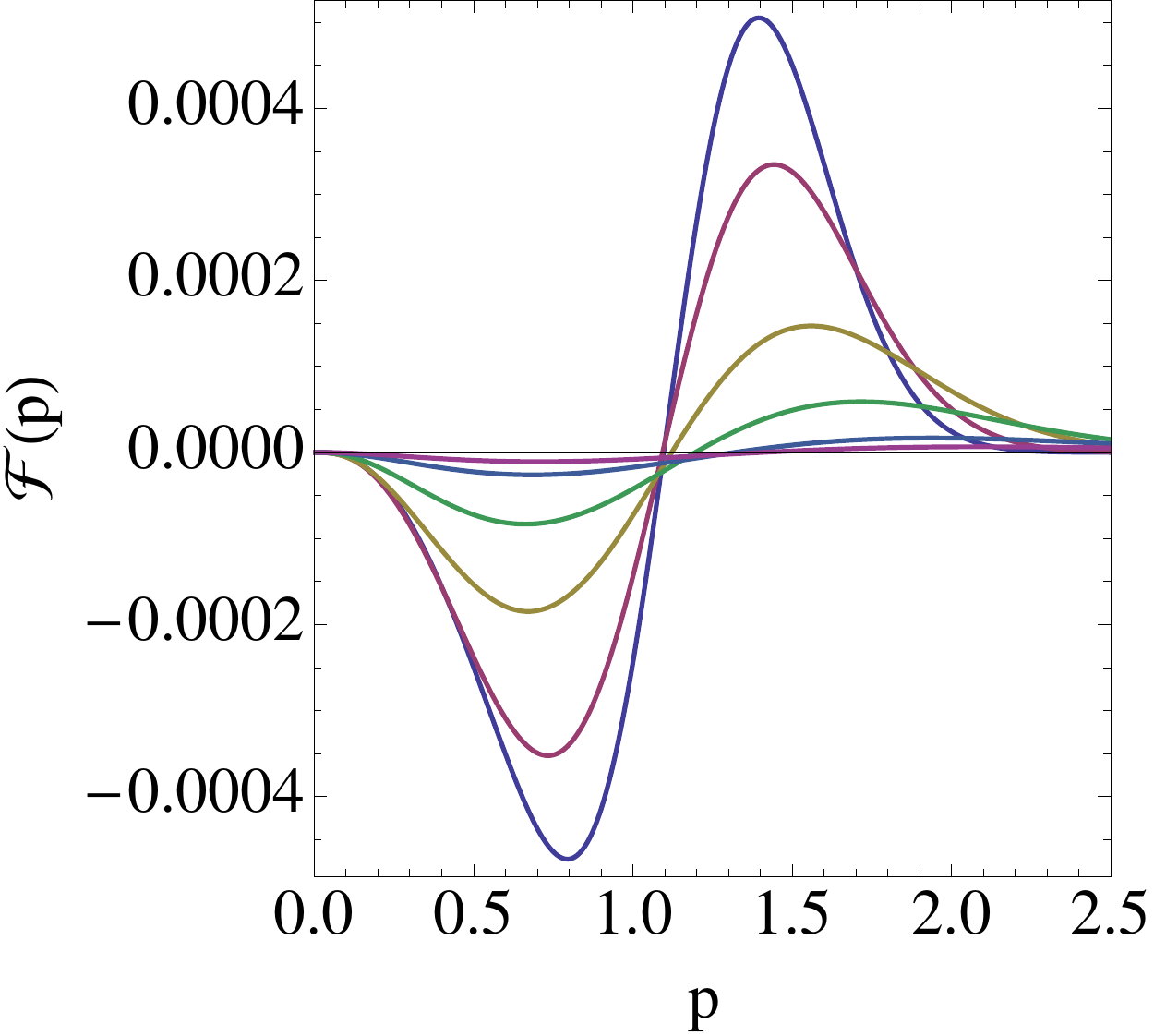} \hspace{0.2cm}
		\includegraphics[width=6cm]{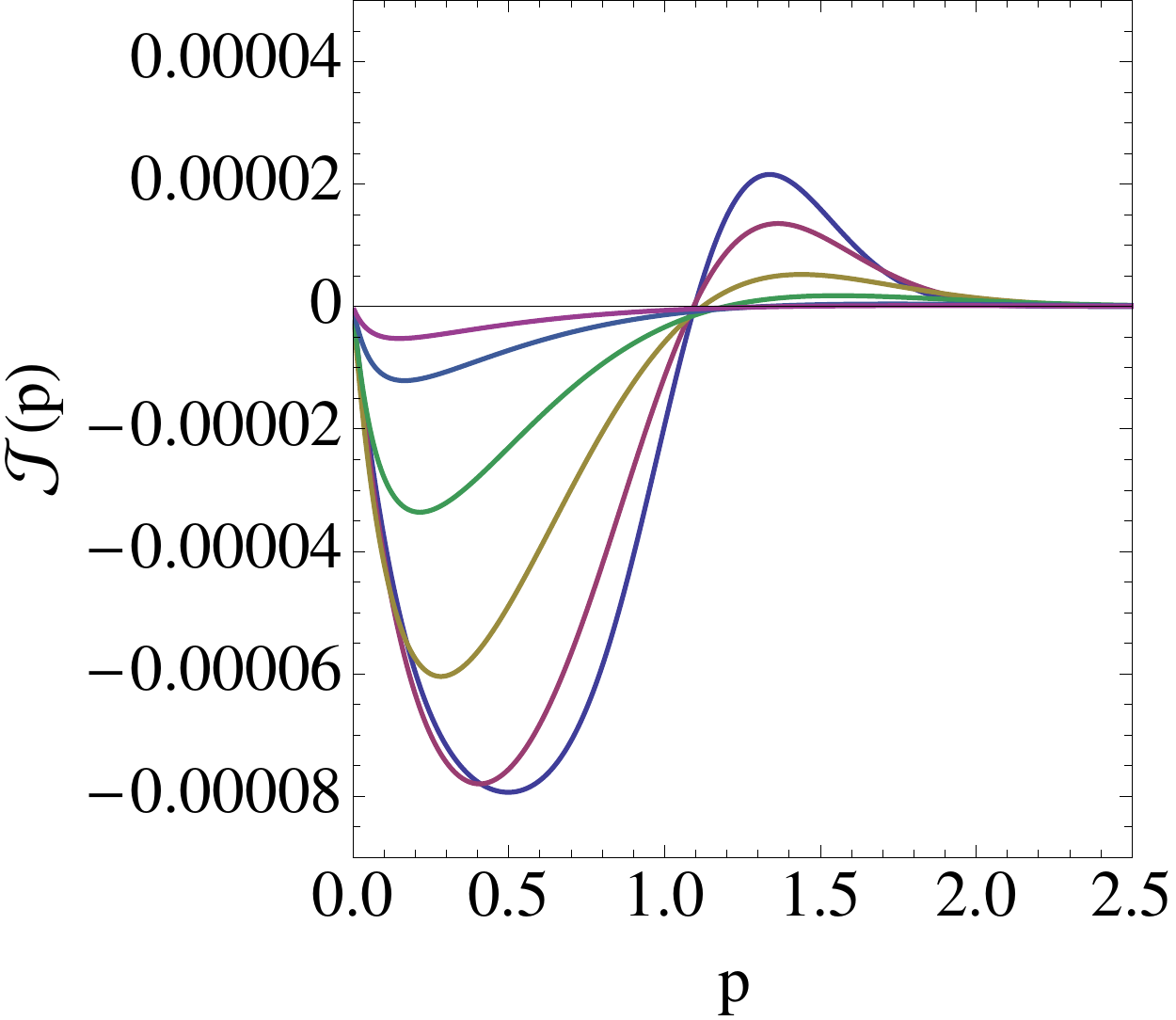}
		\caption{(color online)  The flux ${\mathcal F}(p)$ (left) and the current ${\mathcal J}$ (right) as a function of $p$, for various times between initial time until thermalization is nearly completed  ($f_0=0.1$). The times are $\tau=1.1, 2.7,5.5, 11, 16.5$.}
		\label{fig:fz01_flux}
	\end{center}
%	\vspace{1cm}
\end{figure*}

\section{The over-occupied case: onset of Bose-Einstein condensation}

Now we turn to  the over-populated case, namely the case  $f_0>f_0^c$. For definiteness, most plots presented in this section correspond to $f_0=1$, and CGC initial conditions (except at the end of the section,  where other initial conditions are considered). In this case, the system in equilibrium contains a Bose condensate. As mentioned earlier, we cannot, with the simple equation (\ref{dfdtauJ}), follow the system all the way to equilibrium, but we can study the threshold for the development of the condensate, and show that this occurs in a finite time $\tau_c$. 

\begin{figure*}[t]
	\begin{center}
		\includegraphics[width=6.5cm]{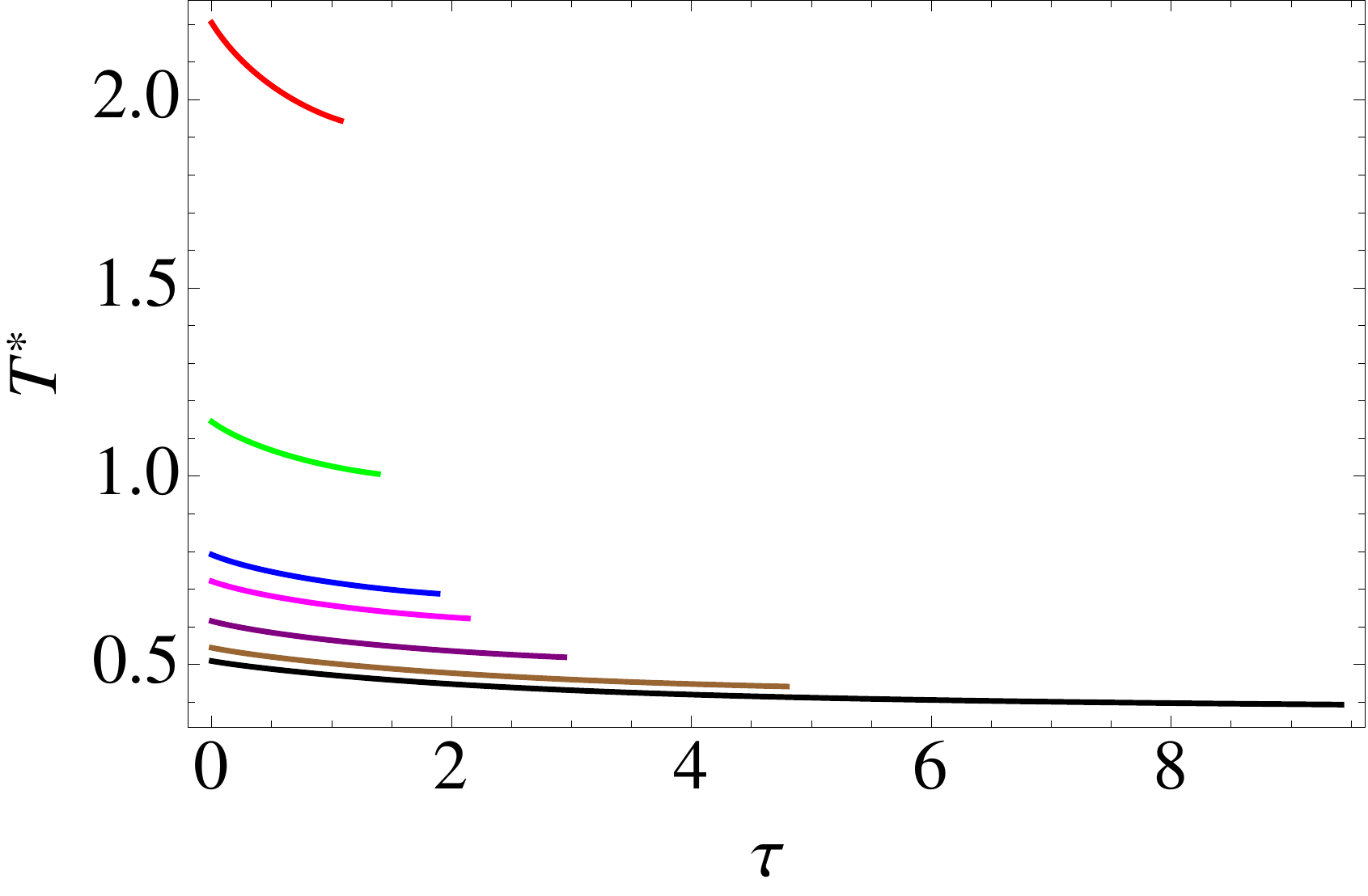} \hspace{0.2cm}
	\includegraphics[width=6.5cm]{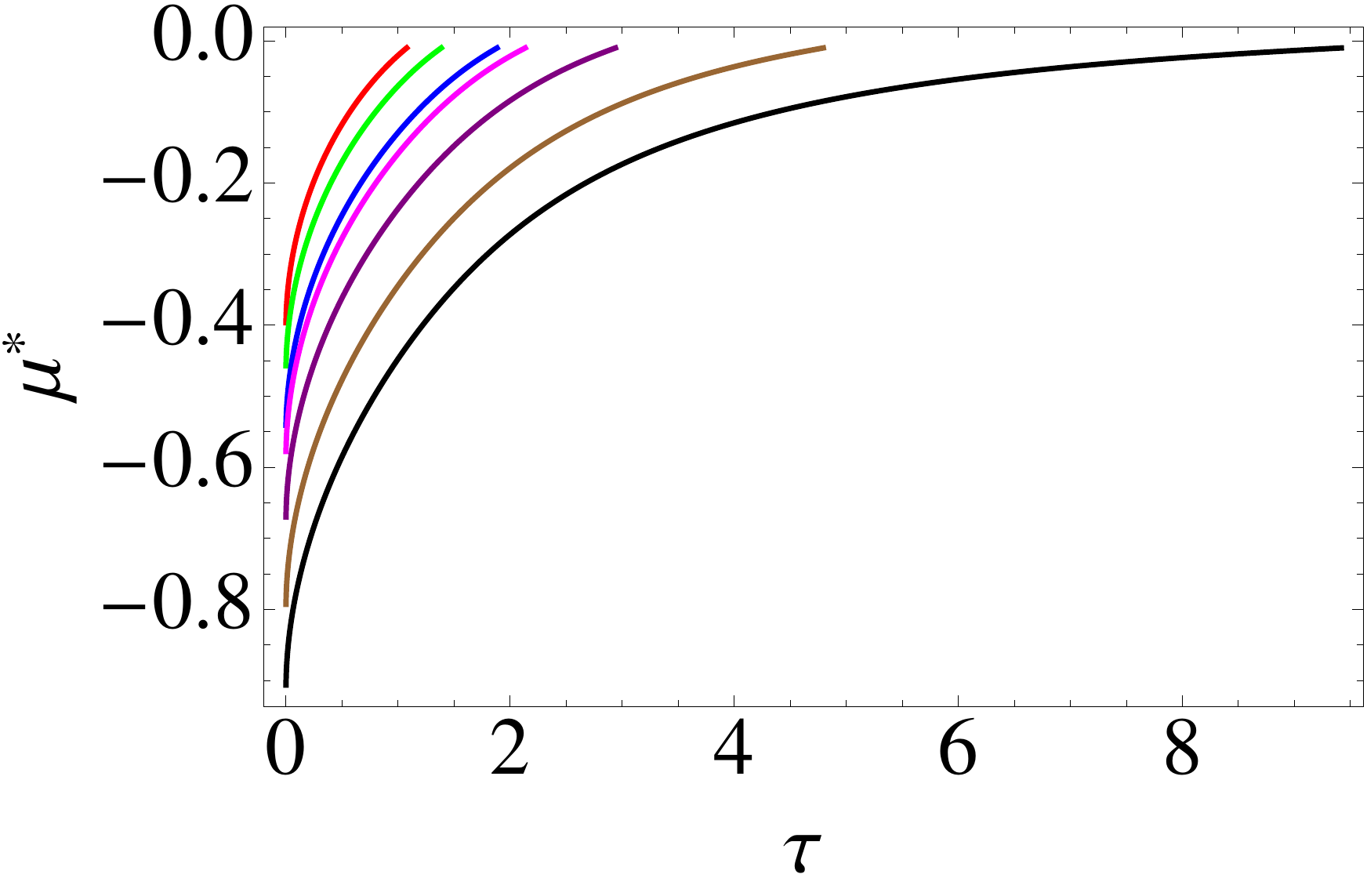}
		\caption{(color online) The effective temperature $T^*$ and chemical potential $\mu^*$ as a function of $\tau$ for various values of $f_0$  (0.2,\,0.3,\,0.5,\,0.8,\,1,\,2,5).   }
		\label{fig:T*f0varl}
	\end{center}
	%\vspace{10cm}
\end{figure*}

A plot of $T^*$ and $\mu^*$ as a function of $\tau$ is given in  Fig.~\ref{fig:T*f0varl}, for various values of $f_0$. The dependence of $T^*$ and $\mu^*$ on $f_0$ at early times can be qualitatively  understood from Eqs.~(\ref{T*0}) which show that for large $f_0$, the initial value of $T^*$  grows linearly with $f_0$, while  $\alpha^*$ decreases with increasing $f_0$, and so does $|\mu^*|$. 

The plot of $\mu^*$ as a function of $\tau$ (right panel of Fig.~\ref{fig:T*f0varl})   clearly shows that $\mu^*$ vanishes for  a finite time $\tau_c$. The value of $\tau_c$ decreases as $f_0$ increases, with $\tau_c$ infinite for $f_0=f_0^c$, as illustrated in the left panel of  Fig.~\ref{fig:tauonset}: since the over-population increases with increasing $f_0$, this says that the larger the over-population, the faster the onset for BEC is reached.  As suggested by the plot in the right panel of Fig.~\ref{fig:T*f0varl}, $\mu^*$ vanishes (approximately) linearly at $\tau_c$. The corresponding slope is  plotted as a function of $f_0$ in the right panel of Fig.~\ref{fig:tauonset}.

\begin{figure*}[h]
	\begin{center}
		\includegraphics[width=6cm]{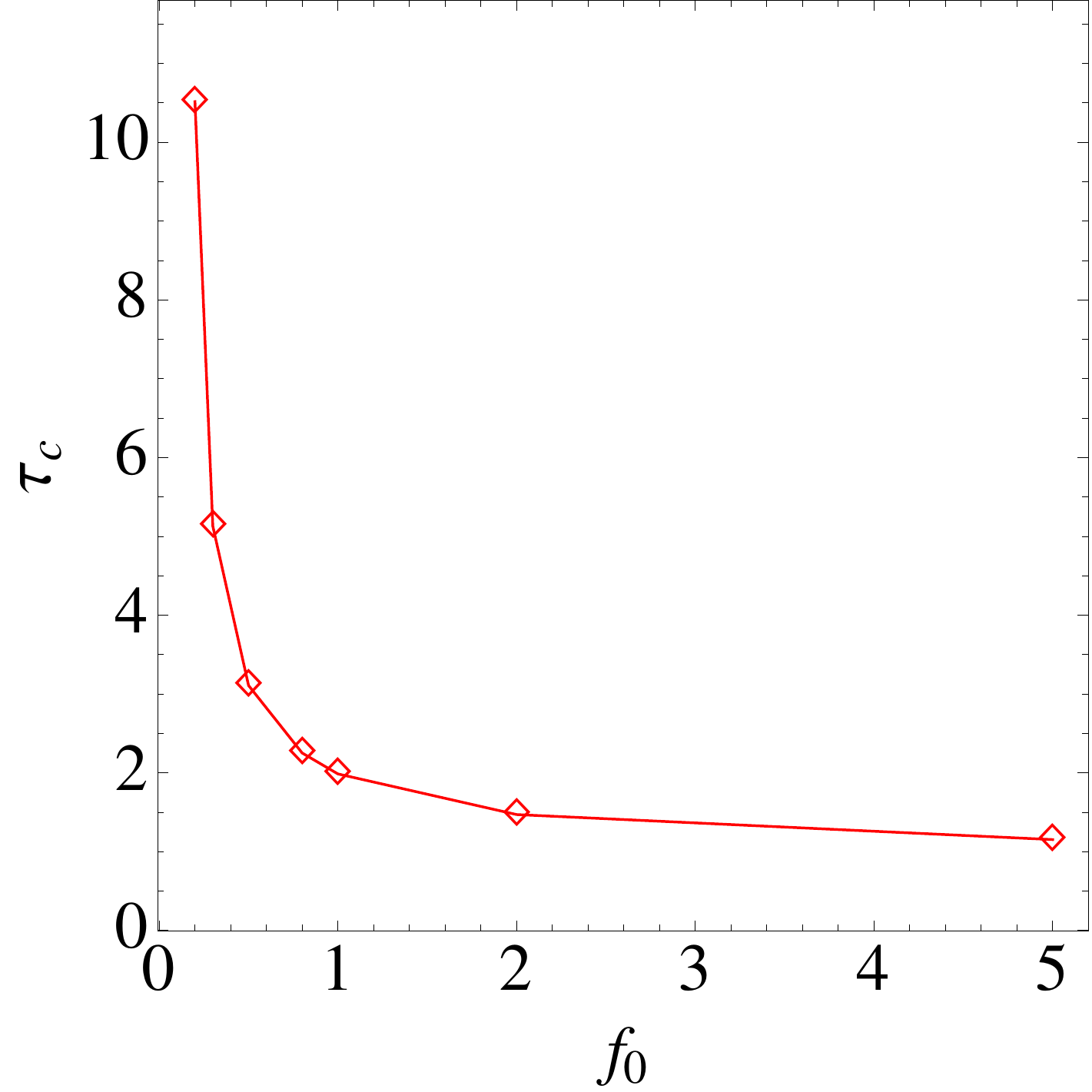}\hspace{0.2cm}
		\includegraphics[width=6cm]{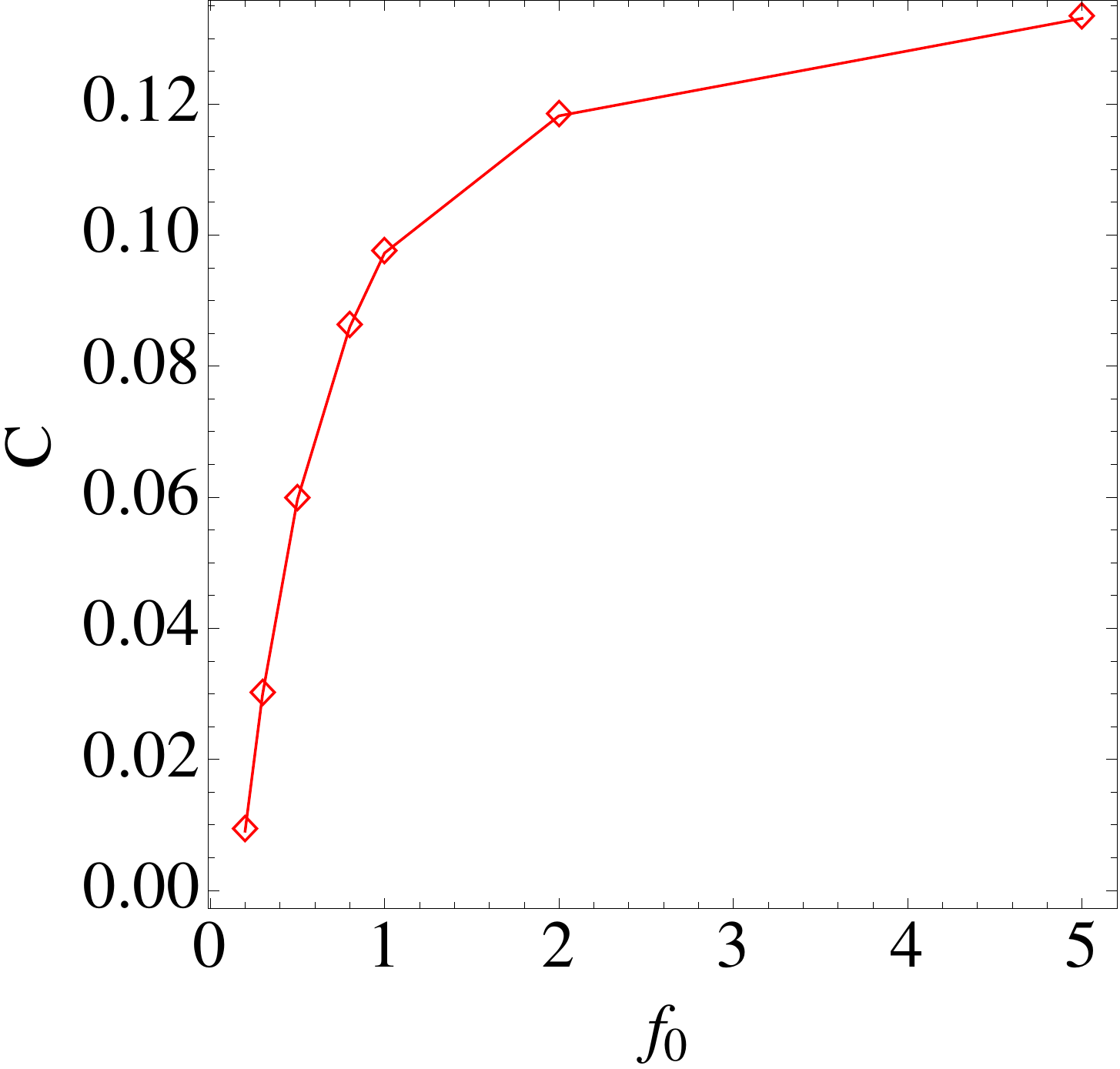}
		\caption{(color online) The value $\tau_c$ at which Bose condensation sets in (left panel). At $\tau_c$, $\mu^*$ vanishes approximately linearly, i.e. $\mu^*\approx C(\tau_c-\tau)$. The slope $C$ is plotted as a function of $f_0$ in the right panel. }
		\label{fig:tauonset}
	\end{center}
	%\vspace{2cm}
\end{figure*}

It is interesting to explore further the limit of large $f_0$. To do so, let us take the distribution  function to be of the form 
 $f(p)=f_0\, g(p/Q_s) $. It is not hard to show that, in the limit $f_0\to\infty$, the kinetic equations (\ref{dfdtauJ0},\ref{calJp0}) reduce to  equations for $g$ from which $f_0$ has been completely eliminated:
 \beq\label{dfdtauJinfty}
\del_\tau g=-\frac{1}{p^2}\frac{\del}{\del p}\left( p^2 \tilde {\mathcal J}(p)  \right),\quad \tilde {\mathcal J}(p)=\!-\!  \partial_p g(p) \int_\q g^2(q) \!-\!  g^2(p) \int_\q \frac{2g(q)}{q}.
\eeq 
The onset of BEC for this equation occurs for at time $\tau_c$ of order 1 ($\tau_c$ is found numerically to be very close to 1 for CGC initial conditions). That there is a non vanishing critical time can be inferred from the following observation: in the limit $f_0\to \infty$, and for CGC initial condition, $|\mu|^*\to Q_s/3$ (see Eq.~(\ref{T*0})). Since it is non vanishing, the system is initially away from the threshold for BEC, and it will take a finite time to reach this threshold. Of course the physical time $t_c$ scales as $\tau_c/f_0^2$, so that as $f_0\to\infty$ the onset of BEC will be reached on an infinitesimal time scale.

Note that the equilibrium is not fully reached at the onset of condensation, so that the notions of temperature and chemical potential are not well defined there. That is, $T^*$ and $\mu^*$ keep their meaning of effective temperature and chemical potential characterizing the small momentum part of the distribution function, as discussed earlier. In order to reach full equilibrium,  the thermal part of the distribution (i.e. that part of the distribution that concern particles not in the condensate) will continue to evolve as condensation proceeds. This tendency is  illustrated in Fig.~\ref{fig:entropyrate} which shows that the rate of the entropy production weakens as one approaches $\tau_c$, but does not vanish at $\tau_c$. Of course, as we have already emphasized, what happens beyond $\tau_c$ can only be studied with a full set of equations that include the coupling between the condensate and the termal particles.

We also wish to stress the crucial role of the Bose statistical factors. With the ``linear'' version of the kinetic equation that was used in Refs.~\cite{Mueller:1999pi} and \cite{Bjoraker:2000cf}, one finds that, for the same over-populated initial condition, the system evolves to a classical Boltzmann distribution with a positive chemical potential, rather than reaching the onset of Bose condensation.

\begin{figure*}[t]
	\begin{center}
	\includegraphics[width=6.5cm]{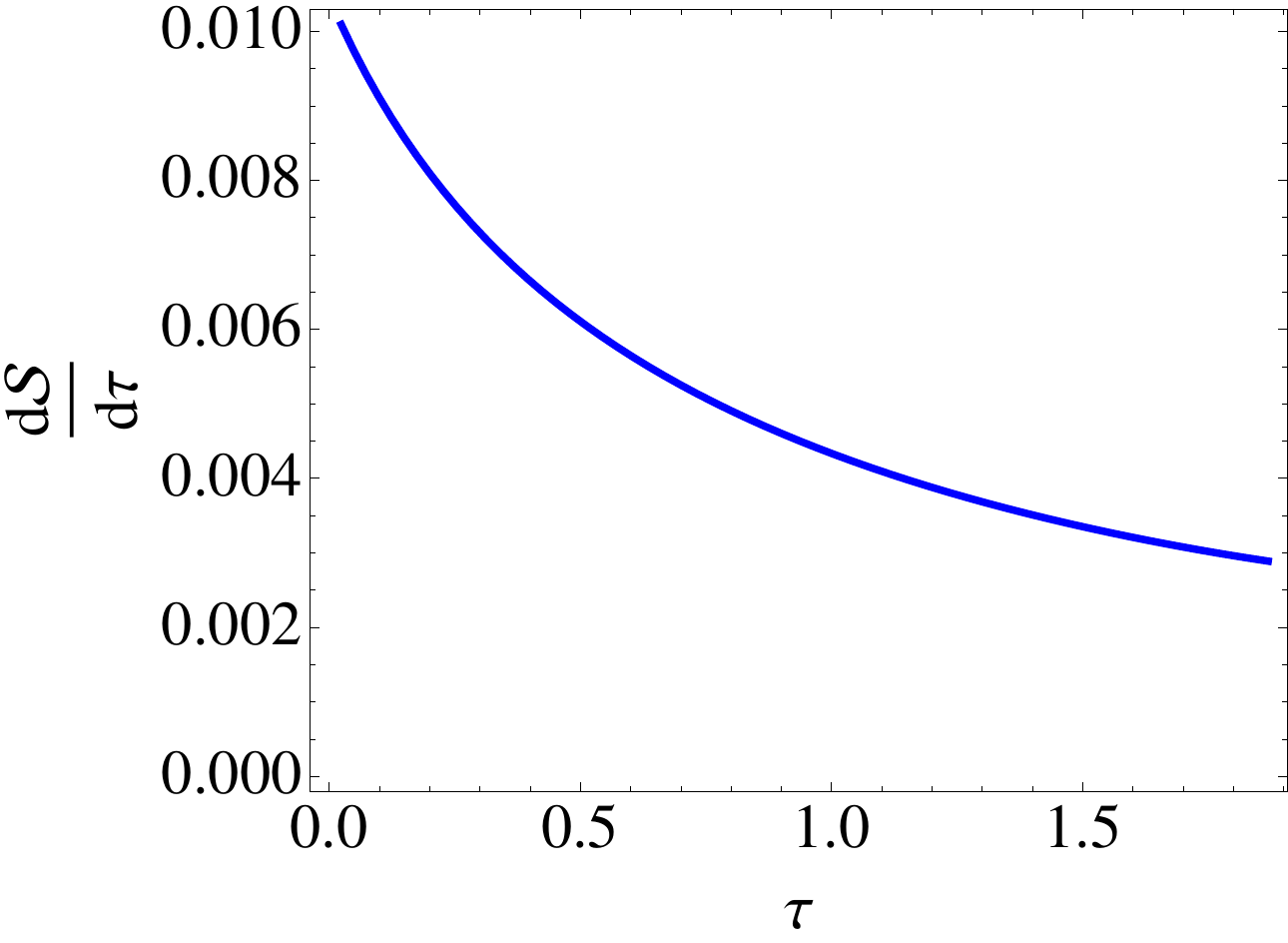}
		\caption{(color online) The rate of change of the entropy, $dS/d\tau$ calculated from Eq.~(\ref{entropyrate}), as a function of $\tau$, for $f_0=1$. }
		\label{fig:entropyrate}
	\end{center}
%	\vspace{-1cm}
\end{figure*}

A more general view of the onset of condensation in terms of the current and flux of particles in momentum space is provided by Fig.~\ref{fig:fz1_Flux}. Particularly noteworthy is the behavior of the current at small momenta: instead of decreasing  smoothly as in the under-populated case, it exhibits a strong increase (in absolute value), suggesting a singular behavior as $\tau$ approaches $\tau_c$. In fact, as we shall see, the low momentum region follows a simple scaling behavior that characterizes the onset of Bose-Einstein condensation. 

\begin{figure*}[h]
	\begin{center}
		\includegraphics[width=6.5cm]{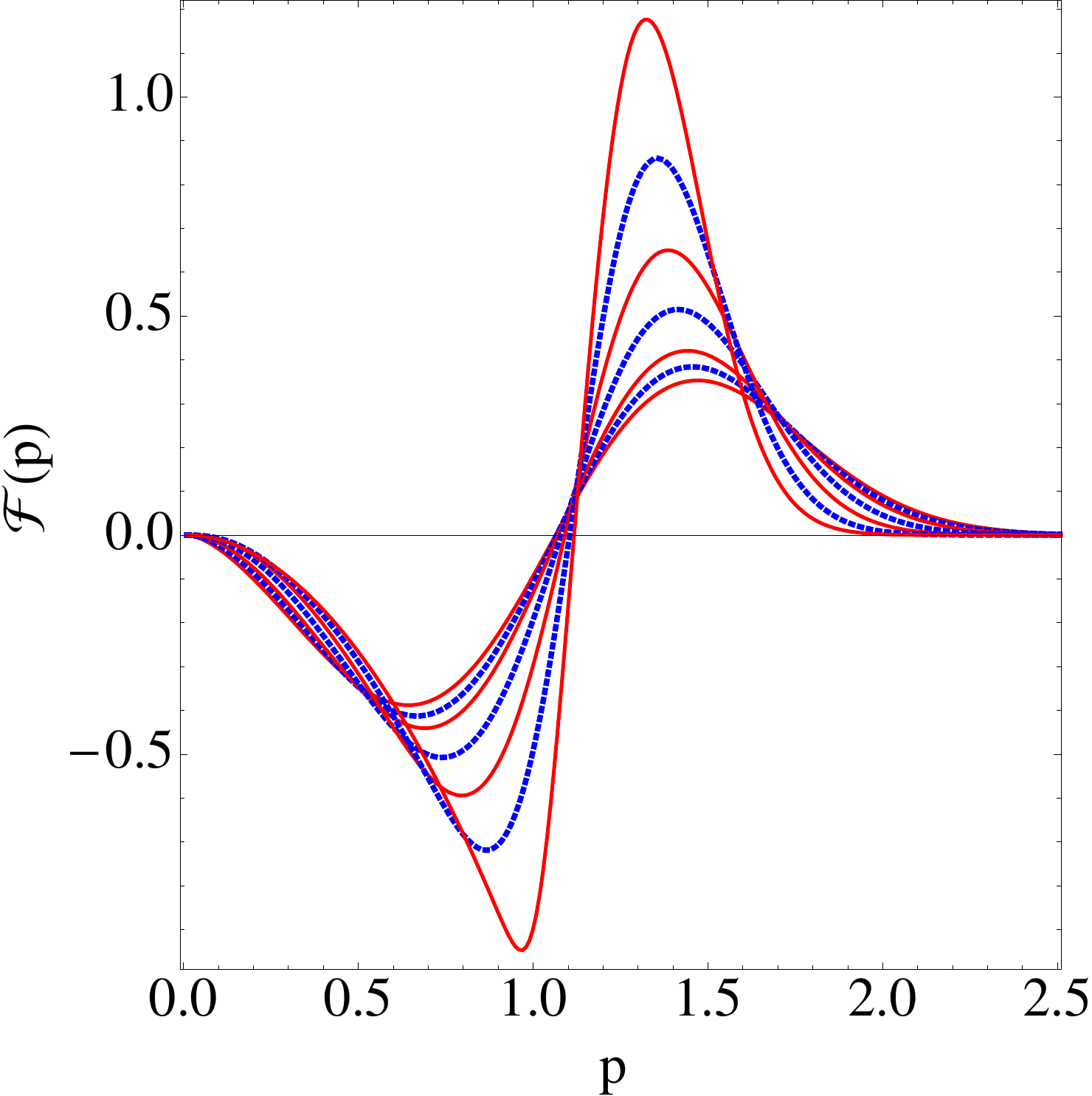} \hspace{0.2cm}
	\includegraphics[width=6.5cm]{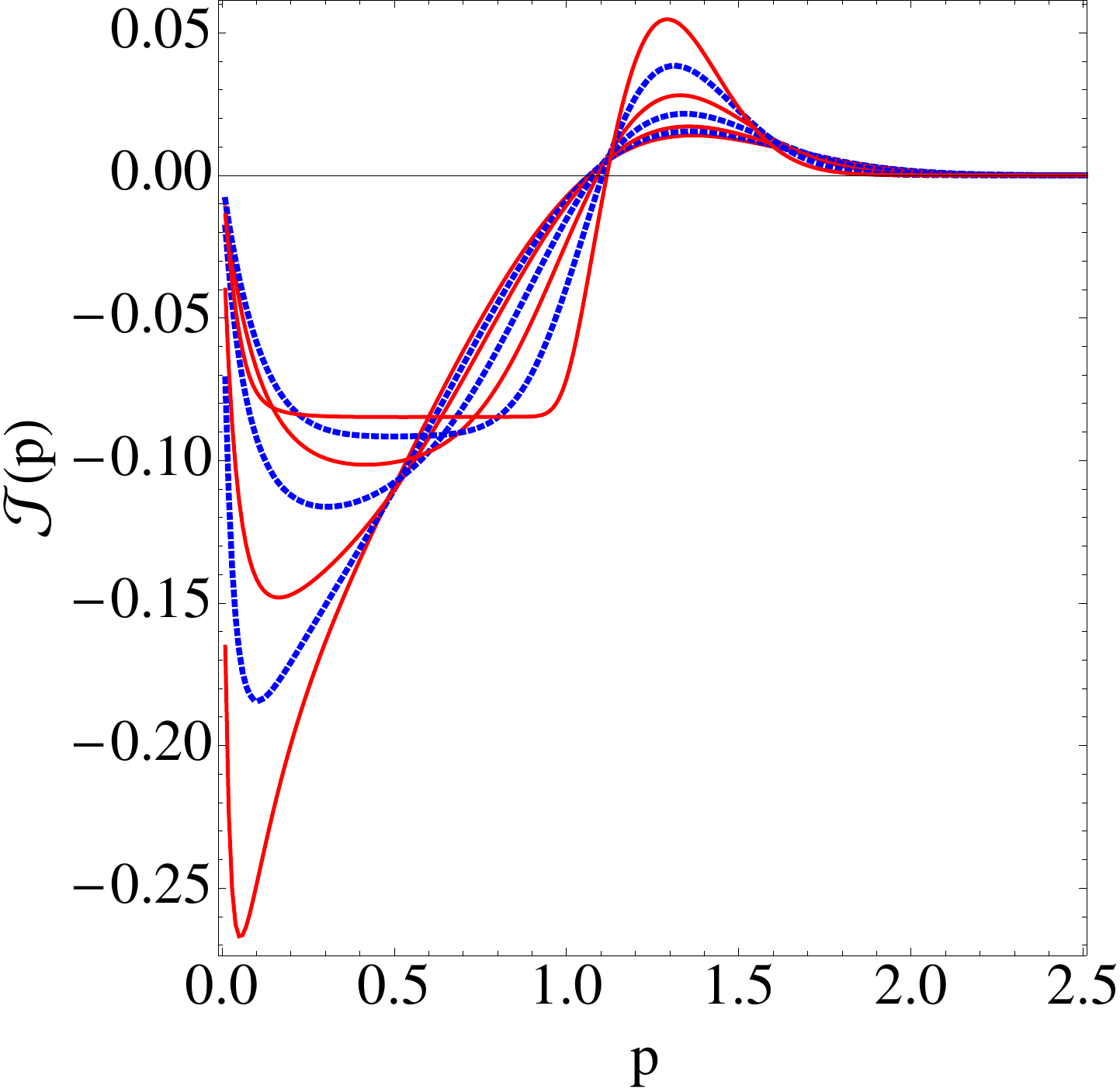}
		\caption{(color online) {The flux ${\mathcal F}(p)$ (left) and the current ${\mathcal J}$ (right) as a function of $p$, for various times between initial time until the time close to onset  ($f_0=1$). The times are $\tau=0.024, 0.25, 0.54, 0.87, 1.23, 1.43, 1.64$.}}
		\label{fig:fz1_Flux}
	\end{center}
%	\vspace{-1cm}
\end{figure*}

In order to understand better how the singularity associated to Bose-Einstein condensation develops as $\tau$  approaches $\tau_c$, we return to the small $p$ analysis, and study the limit $|\mu^*|\to 0$.  In the calculation of the integral $\int_0^{p_0} dp\, p^2 \,\del_\tau f(p)$,  we can approximate the  distribution function by the classical equilibrium distribution, that is $f^*(p)\approx T^*/(p-\mu^*)$, with the parameters $T^*$ and $\mu^*$ function of $\tau$ ($|\mu^*|, p_0\ll T^*$). A simple calculation yields  ($y\equiv p_0/|\mu^*|$):
\beq
\del_\tau \int_0^{p_0} dp\,p^2f(p)= (\mu^*) ^2 \frac{\del T^*}{\del\tau}\,h_1(y)+ T^*|\mu^*|\frac{\del \mu^*}{\del \tau}\,h_2(y),   \eeq
   with
   \beq
   h_1(y)\equiv \log \left(1+y\right)+\frac{1}{2} y (y-2 ),\quad h_2(y)\equiv 2   \log \left(1+y\right)- \frac{y (2 +y)}{1+y}.
   \eeq

This should equal $-p_0^2 {\mathcal J}(p_0)$ (see Eq.~(\ref{smallpeqn})). We have therefore
\beq\label{scalingcurrent}
-{\mathcal J}(\tau, p_0)= \frac{\del T^*}{\del\tau}\frac{h_1(y)}{y^2}+\frac{T^*}{|\mu^*|}\frac{\del \mu^*}{\del \tau} \frac{h_2(y)}{y^2},\quad y\equiv \frac{p_0}{|\mu^*|}.
\eeq

  By expanding the expression above for small $y$, one recovers the expression of the current obtained earlier for $p_0\ll |\mu^*|$, namely
  \beq
  -{\mathcal J}(\tau,p_0)=\left(\frac{1}{|\mu^*|} \frac{\del T^*}{\del\tau}- \frac{T^*}{|\mu^*|^2}\frac{\del |\mu^*|}{\del \tau}\right)p_0 =\frac{\dot f(0)}{3}\,p_0.
  \eeq
  
  Our interest here is the other limit, $|\mu^*|\ll p_0\ll T^*$. In this limit, we find that the current is of the form (in leading order)
\beq
{\mathcal J}(\tau,p_0)\approx \frac{1}{p_0}\, \frac{\del (|\mu^*| T^*)}{\del \tau}.
\eeq
This component of the current is clearly visible in Fig.~\ref{fig:fz1_Flux} (note that a contribution $\sim 1/p$ in the current yields a linear contribution to the flux, also visible in the figure).

\begin{figure*}[h]
	\begin{center}
		\includegraphics[width=6cm]{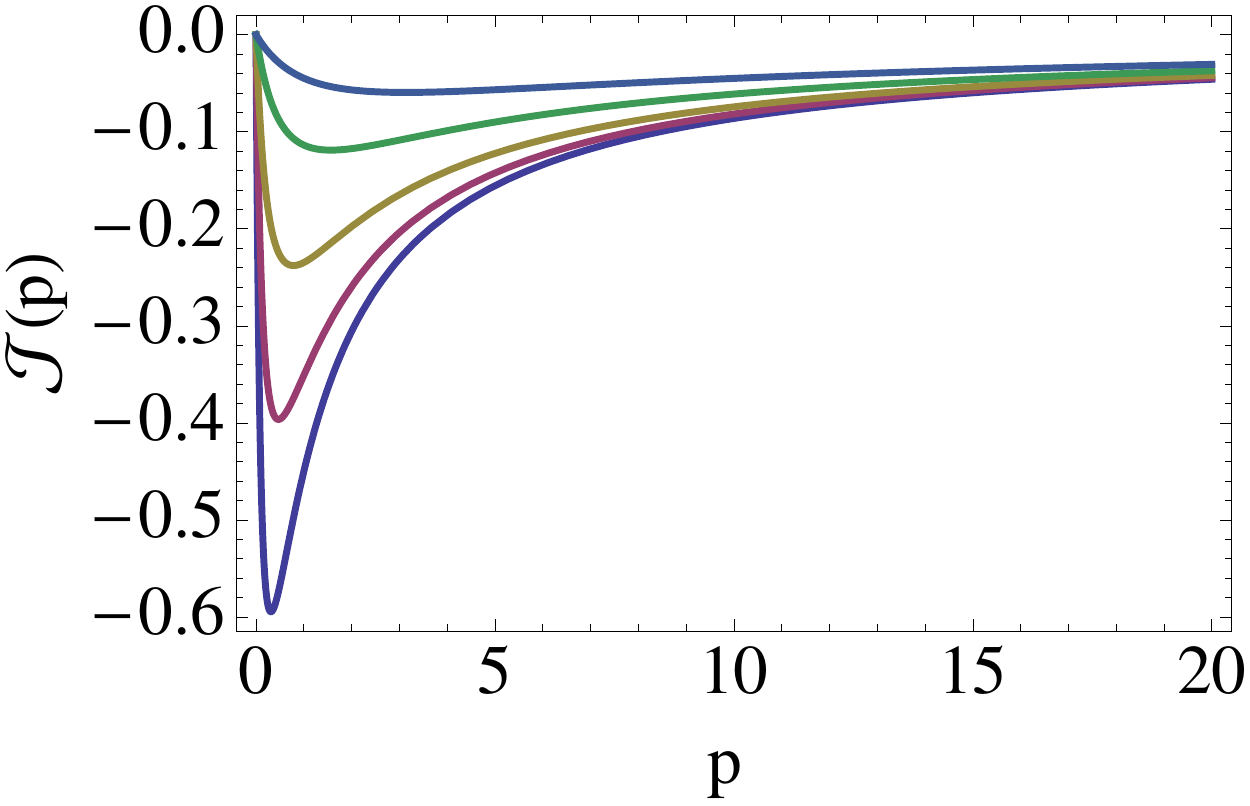} \hspace{0.2cm}
	\includegraphics[width=6cm]{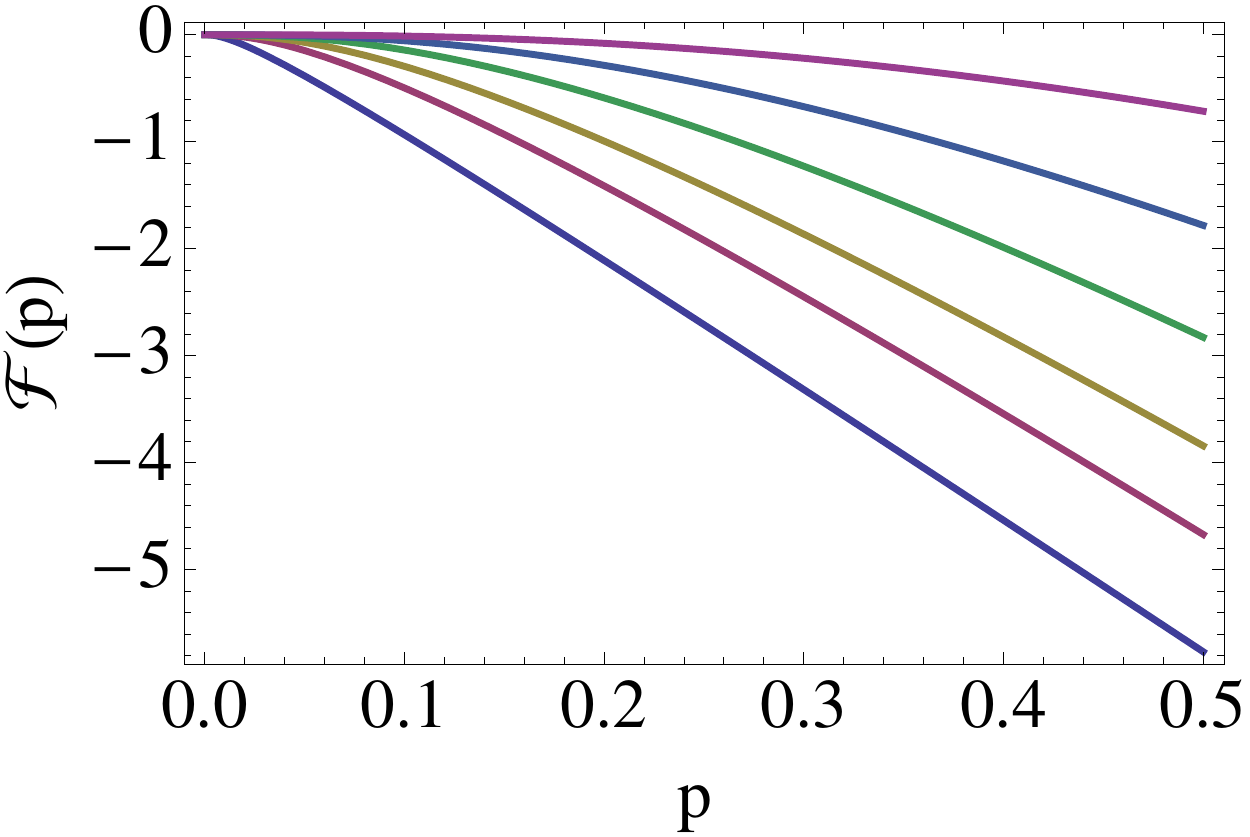}
		\caption{(color online) The scaling function (\ref{scalingcurrent}) representing ${\mathcal J}(p)$ (left) and the corresponding flux ${\mathcal F}(p)$ very near $\tau_c$. }
		\label{fig:Jp_scaling}
	\end{center}
%	\vspace{-1cm}
\end{figure*}

However, more generally, Eq.~(\ref{scalingcurrent}) shows that the current is in a scaling form, with its dependence on $p_0$ entirely contained in the dependence on $y=p_0/|\mu^*|$.  In the vicinity of the critical time, the chemical potential vanishes, and we may set
\beq
|\mu^*|=C(\tau_c-\tau)^\eta.
\eeq
From the discussion above we know that $\eta\simeq 1$.\footnote{In related studies, either in the non relativistic regime \cite{Lacaze:2001qf}, or for massive scalar field theory \cite{Semikoz:1995rd}, where in both cases the dispersion relation is non relativistic in the vicinity of condensation, an exponent larger than 1 is observed.}
As for the temperature, it is a regular function of $\tau$ near $\tau_c$. The scaling function $h_1(y)/y^2$ is an increasing function of $y$, that varies between 0 and 1/2 as $y$ runs from 0 to infinity. In the expression of the current, it multiplies a regular function of $\tau$, ${\del T^*}/{\del\tau}$. The scaling function $h_2(y)/y^2$ is also a regular function of $y$, but it is multiplied by a divergent quantity as $\tau$ approaches $\tau_c$. One concludes then that, at small $p$, and for $\tau$ very near $\tau_c$, the second contribution to the current dominates, so that
\beq
{\mathcal J}(\tau,p)\simeq \frac{\eta T^*}{\tau_c-\tau}\, \frac{h_2(y)}{y^2},\qquad y=\frac{p_0}{C(\tau_c-\tau)^\eta}.
\eeq
This scaling function is plotted in Fig.~\ref{fig:Jp_scaling} (left panel), assuming $\eta=1$, together with the corresponding function representing the flux ${\mathcal F}(p)$ (right panel). One recognizes the characteristic small $p$ behaviors  already identified in Fig.~\ref{fig:fz1_Flux}, in particular  the $1/p$ singularity of the current, and the corresponding linear behavior of the flux. For small values of $y$, $h_2(y)/y^2\sim -y/3$, so that, in terms of $p_0$, the slope of the current diverges at small $p$ at the onset. Also, the function $h_2(y)/y^2$ has a minimum at $y\approx 1.567$. In terms of $p_0$ the corresponding minimum goes to zero as one approaches the onset for condensation. 

\begin{figure*}[h]
	\begin{center}
	\includegraphics[width=12cm]{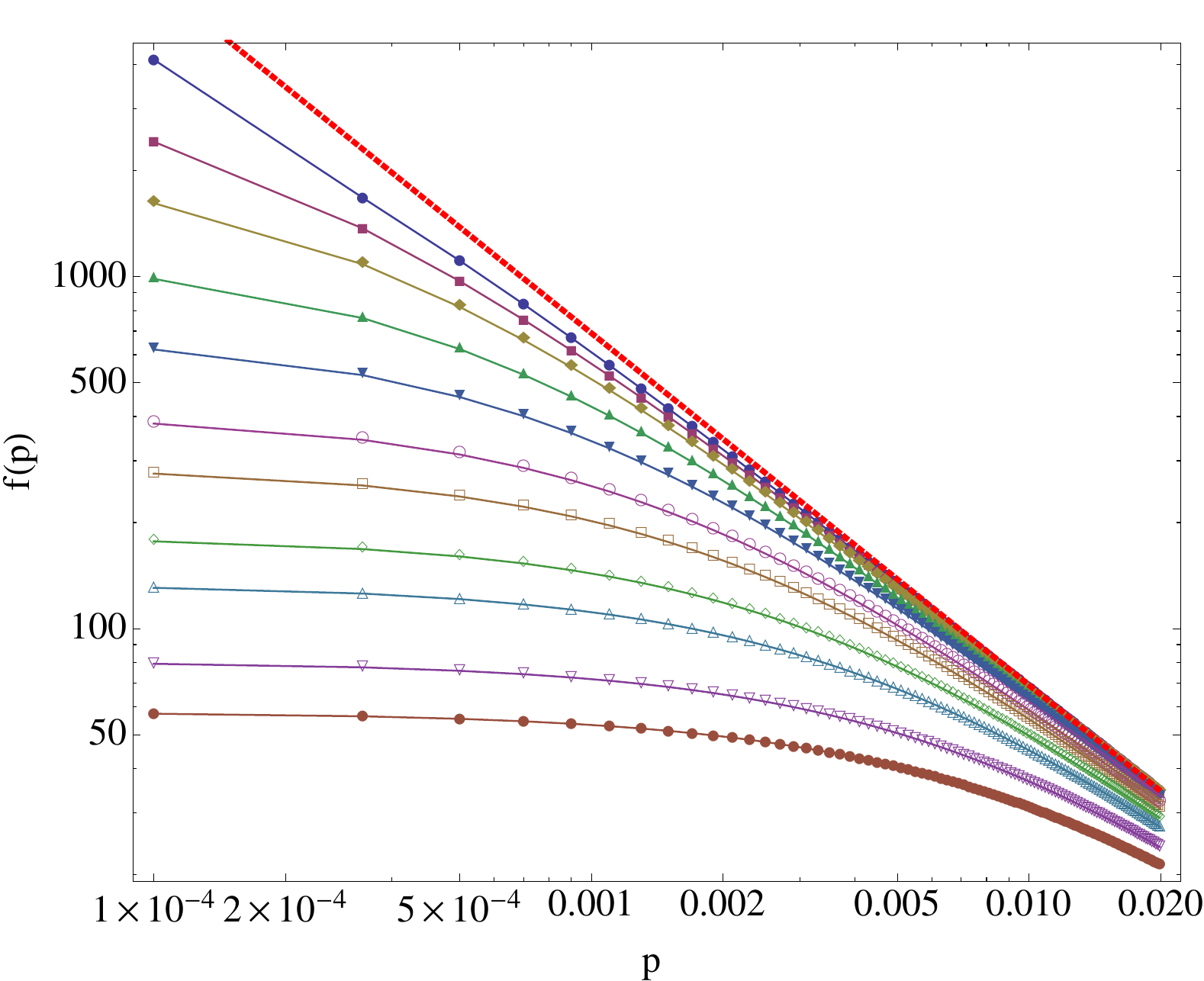}
		\caption{(color online) The distribution  $f(p)$ near the condensation point. The various curves from bottom to top correspond to  $ \tau = 1.8237, 1.8537, 1.8837, 1.8957, 1.9077, 1.9137, 1.9197, 1.9233, 1.9257, 1.9269, 1.9277$. The time at onset is $\tau_c =1.9857$.}
		\label{fig:f_and_current}
	\end{center}
	\vspace{1cm}
\end{figure*}

To complete the picture,  we have plotted in Fig.~\ref{fig:f_and_current} the function  $f(p)$ in a logarithmic scale,  at very small $p$.  This illustrates how the $1/p$ tail of the distribution function develops. At $\tau_c$, the distribution function diverges as $1/p$, but it remains finite at $p=0$ as long as $\mu^*$ is non vanishing.

We conclude this section by presenting results corresponding to different initial conditions. 
As we argued before, we expect the growth of low momentum modes to be a robust feature of the transport equation, for  a large class of relevant initial conditions. As an illustration we have considered a Gaussian initial distribution of the form:
\begin{eqnarray}\label{gaussianIC}
f(p) = f_0 \,g(p/Q_s), \qquad g(p/Q_S)=a\,  e^{-[b*(p/Q_s-1)]^2},
\end{eqnarray}
with the numerical constant $a=0.73$ and $b=4$.  The calculation is done  for  $f_0=1.37$, which corresponds to an over-populated case (for the Gaussian initial condition  (\ref{gaussianIC}) $f_0^c=0.43$).  We expect then the system to reach the onset of Bose-Einstein condensation. And it does.  In contrast to the  color glass initial condition (\ref{finitCGC}), this distribution contains only hard modes to start with, so that the low momentum tail takes more time to build up. However this build up is accelerated by the diffusive part of the current since now the initial distribution presents a positive gradient at small $p$. And indeed, as shown in Fig.~\ref{fig:gaussian}, soft modes quickly appear in the system, and the distribution rapidly acquires a form that is similar to that of the CGC initial condition.  At later times, the characteristic singular universal behavior that signals the proximity of BEC is clearly visible in the plot of the current in the right panel of Fig.~\ref{fig:gaussian}.

\begin{figure*}[h]
	\begin{center}
		\includegraphics[width=5.5cm]{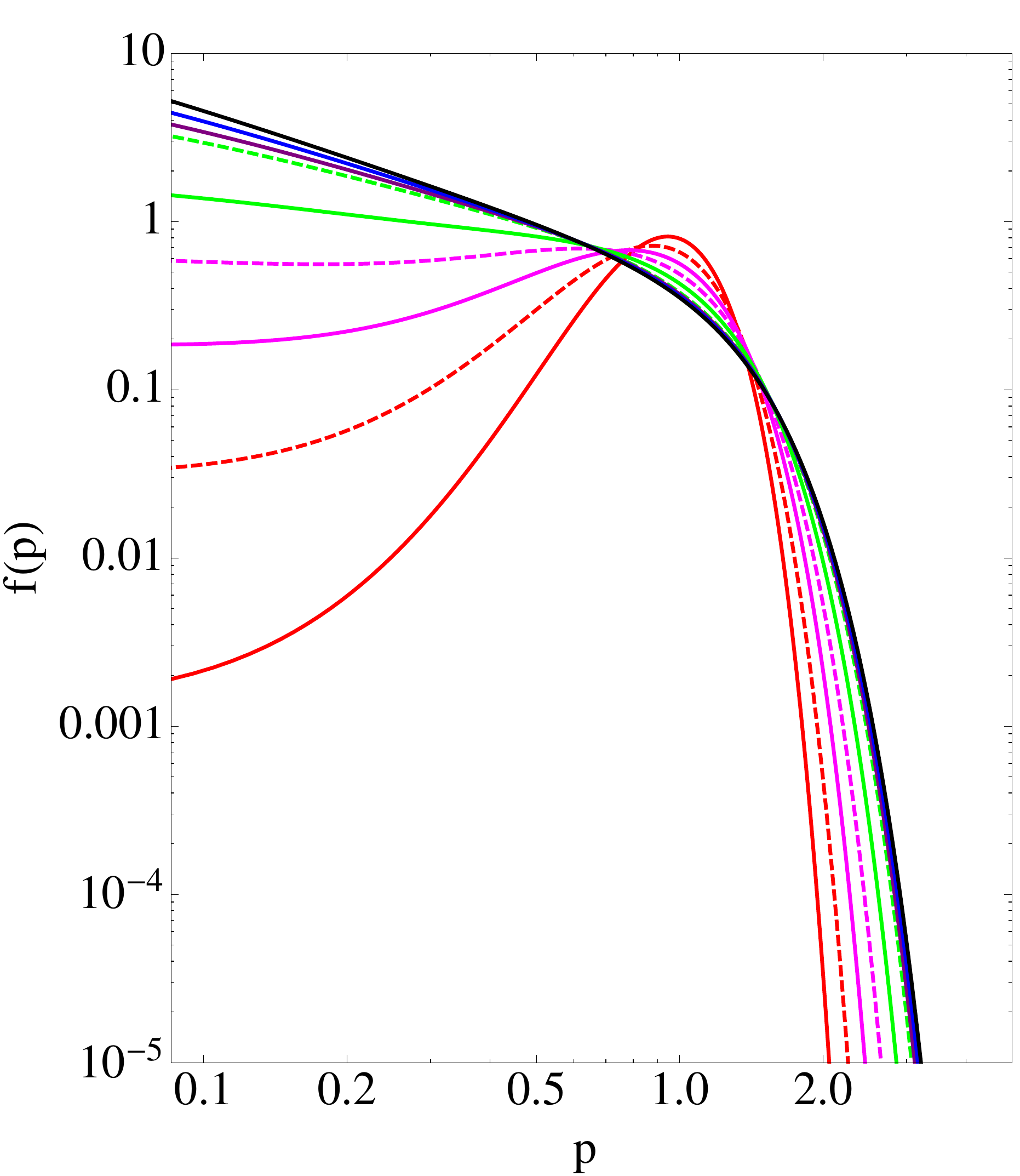}  \hspace{0.2cm}
	\includegraphics[width=5.7cm]{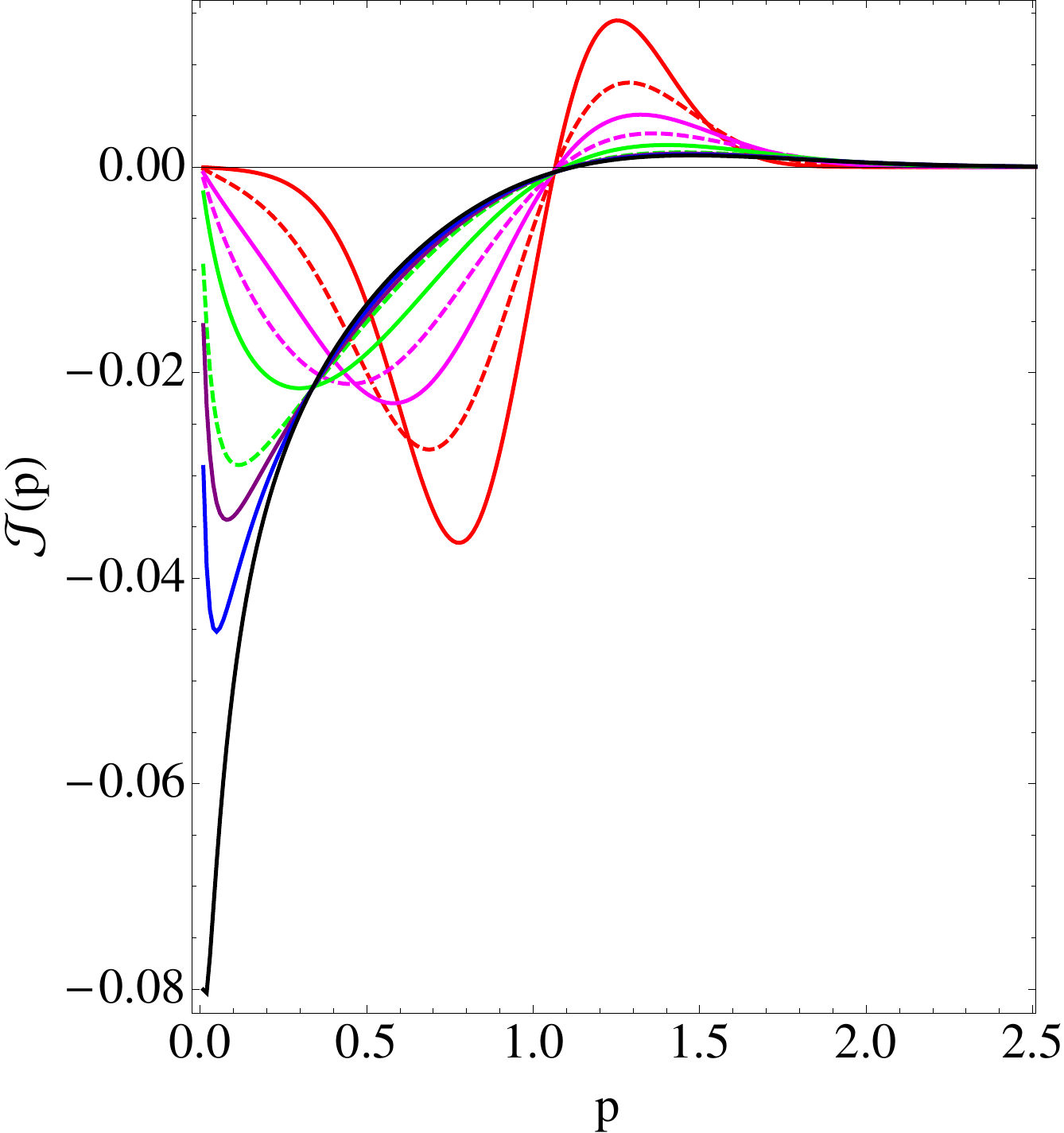}
		\caption{(color online) For the Gaussian initial condition (\ref{gaussianIC}), the distribution function $f(p)$ and the current ${\mathcal J}(p)$ as a function of $p$  for various times, $\tau\approx 0.92, 2.1,3.6,5.6,8.1,11.3,12.1,12.8,13.7$. }
		\label{fig:gaussian}
	\end{center}
		%\vspace{5cm}
\end{figure*}

Finally, although the focus of this paper is elsewhere, it may be interesting to estimate the typical physical time scales involved in the present analysis. Recall that the physical time $t$ and the scaled time $\tau$ are related by (see Eq.~(\ref{taunew}))
\beq\label{physicaltime}
 t = \tau\, \frac{1}{Q_s}\, \frac{1}{f_0(1+f_0)} \, \frac{1}{\alpha_s^2} \, \frac{1}{ 36 \pi {\cal L}}. 
 \eeq
Choosing  $Q_s \sim 2\rm GeV$, and  $\alpha_s^2 \sim 0.1$, one ends up with $ t = \tau /[{f_0(1+f_0)} {\cal L}]$, with $\tau$ dimensionless and $t$ in fm/c.  Assuming ${\cal L}$ to be of order unity, the coefficient of $\tau$ in Eq.~(\ref{physicaltime}) is a number of order unity for generic values of $f_0$. It follows that $t$ and $\tau$ are  of the same order of magnitude. The typical time scales  involved in our analysis range therefore typically from 1 fm/c (overpopulation) to 10 fm/c (underpopulation). Much shorter time scales appear in case of large overpopulation. Thus, if $f_0\sim 1/\alpha$, we have instead $t = 0.1\, \tau /{\cal L}$, in the 0.1 fm/c range. All these numbers are of the same order of magnitudes as the typical time scales expected in the early stages of heavy ion collisions.

\section{Conclusion}

In this paper, we have provided numerical evidence that  a system of gluons with an initial distribution that mimic that expected in heavy ion collisions reaches the onset of Bose-Einstein condensation in a finite time. This was obtained by using  kinetic theory, with a quantum Boltzmann equation in the small angle approximation, which reduces then to a Fokker-Planck equation that can be solved with elementary numerical techniques.  The role of  Bose statistical factors in amplifying the rapid  growth of the population of the soft modes has been emphasized. With these factors properly taken into account, one finds that elastic scattering alone provides an efficient mechanism for populating soft modes, that could be competitive with the radiation mechanism invoked in the scenario of Ref.~\cite{Baier:2000sb}.

The present study suffers from a number of obvious limitations that  need to be addressed before any conclusion can be reached regarding the relevance of the phenomena discussed here to the thermalization of the quark-gluon plasma. Natural extensions of the present work are being studied. These include  the effect of the  longitudinal expansion,  the evolution of the system after the onset of condensation, the (approximate) treatment of inelastic scatterings. We shall report on these issues in forthcoming publications.

\section*{Acknowledgements}
The research of  LM and JL is supported under DOE Contract No. DE-AC02-98CH10886. JL is also grateful to the RIKEN BNL Research Center for partial support. The research of JPB is supported by the European Research Council under the Advanced Investigator Grant ERC-AD-267258.


\begin{thebibliography}{00}

 %\cite{Blaizot:2011xf}
\bibitem{Blaizot:2011xf}
  J.~-P.~Blaizot, F.~Gelis, J.~Liao, L.~McLerran, R.~Venugopalan,
  %``Bose--Einstein Condensation and Thermalization of the Quark Gluon Plasma,''
Nucl.\ Phys.\ {\bf A873}, 68-80 (2012). [arXiv:1107.5296[hep-ph]].

%\cite{Mueller:1999pi}
\bibitem{Mueller:1999pi}
 A.~H.~Mueller,
  %``The Boltzmann equation for gluons at early times after a heavy ion
  %collision,''
  Phys.\ Lett.\  B {\bf 475}, 220 (2000)
  [arXiv:hep-ph/9909388].
  %%CITATION = PHLTA,B475,220;%%


%\cite{Bjoraker:2000cf}
\bibitem{Bjoraker:2000cf}
  J.~Bjoraker and R.~Venugopalan,
  %``From colored glass condensate to gluon plasma: Equilibration in high
  %energy heavy ion collisions,''
  Phys.\ Rev.\  C {\bf 63}, 024609 (2001)
  [arXiv:hep-ph/0008294].
  %%CITATION = PHRVA,C63,024609;%%
  
  %\cite{Huang:2013lia}
\bibitem{Huang:2013lia}
  X.~-G.~Huang and J.~Liao,
  %``Glasma Evolution and Bose Condensation with Elastic and Inelastic Collisions,''
  arXiv:1303.7214 [nucl-th].
  %%CITATION = ARXIV:1303.7214;%%



%\cite{Semikoz:1995rd}
\bibitem{Semikoz:1995rd}
  D.~V.~Semikoz and I.~I.~Tkachev,
  %``Condensation of bosons in kinetic regime,''
  Phys.\ Rev.\ D {\bf 55} (1997) 489
  [hep-ph/9507306].
  %%CITATION = HEP-PH/9507306;%%
  %22 citations counted in INSPIRE as of 23 Apr 2013


%\cite{Berges:2012us}
\bibitem{Berges:2012us}
  J.~Berges and D.~Sexty,
  %``Bose condensation far from equilibrium,''
  Phys.\ Rev.\ Lett.\  {\bf 108} (2012) 161601
  [arXiv:1201.0687 [hep-ph]].
  %%CITATION = ARXIV:1201.0687;%%
  %19 citations counted in INSPIRE as of 26 Apr 2013
  
  %\cite{Dusling:2010rm}
\bibitem{Dusling:2010rm}
  K.~Dusling, T.~Epelbaum, F.~Gelis and R.~Venugopalan,
  %``Role of quantum fluctuations in a system with strong fields: Onset of hydrodynamical flow,''
  Nucl.\ Phys.\ A {\bf 850} (2011) 69
  [arXiv:1009.4363 [hep-ph]].
  %%CITATION = ARXIV:1009.4363;%%
  %38 citations counted in INSPIRE as of 26 Apr 2013

%\cite{Berges:2013eia}
\bibitem{Berges:2013eia}
  J.~Berges, K.~Boguslavski, S.~Schlichting and R.~Venugopalan,
  %``Turbulent thermalization of the Quark Gluon Plasma,''
  arXiv:1303.5650 [hep-ph].
  %%CITATION = ARXIV:1303.5650;%%
  %3 citations counted in INSPIRE as of 26 Apr 2013
  
  %\cite{Berges:2012ks}
\bibitem{Berges:2012ks}
  J.~Berges, J.~-P.~Blaizot and F.~Gelis,
  %``EMMI Rapid Reaction Task Force on 'Thermalization in Non-abelian Plasmas',''
  J.\ Phys.\ G {\bf 39} (2012) 085115
  [arXiv:1203.2042 [hep-ph]].
  %%CITATION = ARXIV:1203.2042;%%



%\cite{Kurkela:2011ub}
\bibitem{Kurkela:2011ub}
  A.~Kurkela and G.~D.~Moore,
  %``Bjorken Flow, Plasma Instabilities, and Thermalization,''
  JHEP {\bf 1111} (2011) 120
  [arXiv:1108.4684 [hep-ph]]; 
  %%CITATION = ARXIV:1108.4684;%%
  %26 citations counted in INSPIRE as of 26 Apr 2013
%\cite{Kurkela:2011ti}
%\bibitem{Kurkela:2011ti}
 % A.~Kurkela and G.~D.~Moore,
  %``Thermalization in Weakly Coupled Nonabelian Plasmas,''
  JHEP {\bf 1112} (2011) 044
  [arXiv:1107.5050 [hep-ph]].
  %%CITATION = ARXIV:1107.5050;%%
  %33 citations counted in INSPIRE as of 26 Apr 2013

  
  %\cite{Lacaze:2001qf}
\bibitem{Lacaze:2001qf}
  R.~Lacaze, P.~Lallemand, Y.~Pomeau and S.~Rica,
  %``Dynamical formation of a Bose-Einstein condensate,''
  Physica D {\bf 152} (2001) 779.
  %%CITATION = PHYSA,D152,779;%%
  %2 citations counted in INSPIRE as of 26 Apr 2013





 %\cite{Peskin:1995ev}
\bibitem{Peskin:1995ev}
  M.~E.~Peskin, D.~V.~Schroeder,
``An Introduction to quantum field theory,''
  Reading, USA: Addison-Wesley (1995) 842 p.


\bibitem{LifshitzPitaevskii}
E.M. Lifshitz and L.P. Pitaevskii, Physical Kinetics (Pergamon Press, New York, 1981).






%\cite{Baier:2000sb}
\bibitem{Baier:2000sb}
  R.~Baier, A.~H.~Mueller, D.~Schiff and D.~T.~Son,
  %``'Bottom-up' thermalization in heavy ion collisions,''
  Phys.\ Lett.\  B {\bf 502}, 51 (2001)
  [arXiv:hep-ph/0009237].
  %%CITATION = PHLTA,B502,51;%%






\end{thebibliography}
\end{document}